\pdfoutput=1 
\documentclass[a4paper,12pt]{article}

\usepackage[bbgreekl]{mathbbol}
\usepackage{mathrsfs}
\usepackage{graphicx}
\usepackage{amsmath}
\numberwithin{figure}{section}
\usepackage{amsfonts}
\usepackage{indentfirst}
\usepackage{amsthm}
\usepackage{amssymb}
\usepackage{xcolor}
\usepackage{geometry}
\usepackage{url}
\usepackage{setspace}
\usepackage{verbatim}
\usepackage{ulem}
\usepackage{multirow}
\usepackage{booktabs}
\usepackage[page,toc,titletoc,title]{appendix}
\usepackage[noend]{algpseudocode} 
\usepackage{algorithm}
\usepackage{subfigure}
\usepackage{caption}

\numberwithin{equation}{section} 
\newtheorem{theorem}{Theorem}[section]

\newtheorem{definition}{Definition}[section]

\newcommand{\Vc}{\mathcal{V}}

\newcommand{\Fc}{\mathcal{F}}
   
\newcommand{\R}{\mathbb{R}}
\newcommand{\RL}{\mathcal{R}}

\newcommand{\Sc}{\mathcal{S}}
\newcommand{\Oc}{\mathcal{O}}
\newcommand{\C}{\mathbb{C}}

\newcommand{\Z}{\mathbb{Z}}

\newcommand{\Ec}{E_{\rm c}}

\newcommand{\dd}{~{\rm d}}

\newcommand{\vecn}[1]{{\Phi}^{#1}}   
\newcommand{\vm}[1]{\hat{v}^{(1)}_{#1}} 
\newcommand{\vn}[1]{\hat{v}^{(2)}_{#1}} 
\newcommand{\vj}[1]{\hat{v}^{(j)}_{#1}} 
\newcommand{\Index}[1]{\mathcal{I}^{#1}_{\Ec}} 
\newcommand{\xqedhere}[2]{%
\rlap{\hbox to#1{\hfil\llap{\ensuremath{#2}}}}}

\newcommand\tbbint{{-\mkern -16mu\int}}

\newcommand\dbbint{{-\mkern -19mu\int}}

\newcommand\intRd{
{\mathchoice{\dbbint}{\tbbint}{\tbbint}{\tbbint}}
}

\title{Layer-splitting methods for time-dependent Schr\"{o}dinger equations of incommensurate systems
}
\author{
Ting Wang\thanks{
{\it wangting2019@mail.bnu.edu.cn},
School of Mathematical Sciences, Beijing Normal University, Beijing 100875, China.
},
~Huajie Chen\thanks{
{\it chen.huajie@bnu.edu.cn},
School of Mathematical Sciences, Beijing Normal University, Beijing 100875, China.
},
~Aihui Zhou\thanks{
{\it azhou@lsec.cc.ac.cn},
LSEC, Institute of Computational Mathematics and Scientific/Engineering Computing, Academy of Mathematics and Systems Science, Chinese Academy of Sciences, Beijing 100190, China;
School of Mathematical Sciences, University of Chinese Academy of Sciences, Beijing 100049, China.
}
~and~
Yuzhi Zhou\thanks{
{\it zhou\_yuzhi@iapcm.ac.cn},
CAEP Software Center for High Performance Numerical Simulation, Beijing 100088, China;
Institute of Applied Physics and Computational Mathematics, Beijing 100094, China.
}
}
\date{}

\begin{document}

\maketitle

\begin{abstract}
This work considers numerical methods for the time-dependent Schr\"{o}dinger equation of incommensurate systems.
By using a plane wave method for spatial discretization, the incommensurate problem is lifted to a higher dimension that results in semidiscrete differential equations with extremely demanding computational cost.
We propose several fully discrete time stepping schemes based on the idea of ``layer-splitting", which decompose the semidiscrete problem into sub-problems that each corresponds to one of the periodic layers.
Then these schemes handle only some periodic systems in the original lower dimension at each time step, which reduces the computational cost significantly and is natural to involve stochastic methods and parallel computing.
Both theoretical analysis and numerical experiments are provided to support the reliability and efficiency of the algorithms.

\end{abstract}

\tableofcontents

\section{Introduction}
\label{sec:introduction}

Recently there have been extensive research efforts on low dimensional materials with multilayer structures \cite{cao2018unconventional,geim2013van,mele2012interlayer,novoselov20162d, rozhkov2016electronic}.
The mechanical and electronic properties of such systems depend heavily on the stacking arrangement, which allows various manipulation to create heterostructure devices \cite{britnell2012field,ponomarenko2011tunable}.
The lattice periods of the individual layers are generally incommensurate due to the difference in the crystal structure and also due to misorientation between the adjacent layers \cite{dai2016twisted,pereira2006lopes,sboychakov2015electronic}.
Meanwhile, in the experiments of ultracold atoms \cite{deissler2010delocalization,roati2008anderson} and photonic crystals \cite{segev2013anderson,wang2020localization} the incommensurate potentials are created by the interference of laser or the modulation of the refractive indeces and people have observed the localized-extended transition in the time evolution of the quantum waves.

The absence of periodicity in the incommensurate systems presents an essential challenge to atomic/electronic structure calculations since the Bloch theorem can not be applied directly. 
The conventional method to study such systems is to strain them to some commensurate supercells such that the periodicity can be retained 
\cite{koda2016coincidence,komsa2013electronic,loh2015graphene,terrones2014bilayers}. 
However, this type of approaches are usually very expensive and there is no verification of the approximation error. 
Recently, a plane wave method has been proposed to study the quasicrystals \cite{jiang2014numerical} and eigenvalue problems of the incommensurate systems \cite{chen2020plane,zhou2019plane}, which in principle gets rid of the modeling error and is efficient in numerical simulations.
Despite many advantages of the plane wave method for incommensurate systems, it has been shown in \cite{chen2020plane,zhou2019plane} that this method essentially lifts the problem into a higher dimensional space and therefore results in a discrete problem with huge degrees of freedom, especially when the system consists of many periodic layers and when a large energy cutoff is used for the plane wave discretization.
This difficulty needs to be resolved so that one can achieve ``affordable" simulations with high accuracy for incommensurate systems, in particular, for the time-dependent problems.

The goal of this paper is to apply the plane wave method to the time-dependent Schr\"{o}dinger equations of incommensurate systems for spacial discretization, and then design efficient time-stepping schemes to cure the problem of dimension lifting.
Our idea is to split the total Hamiltonian operator such that each part is related to one of the periodic layers in the incommensurate system.
Based on this ``layer-splitting" idea, semidiscrete problem is split into low dimensional sub-problems, and the wavefunctions are evolving by several periodic operators at each time step rather than an incommensurate one,
which can reduce the computational cost significantly (see more discussions in Section \ref{sec:temperoral discretization}).
The layer-splitting methods are natural to involve parallel computing and stochastic algorithms, thus very beneficial for simulating large-scale systems.

The algorithms developed in this paper are analogous to the so-called operator splitting methods \cite{glowinski2017splitting,marchuk1968some,strang1968construction} that have been widely used in the numerical solutions of partial differential equations, especially for the time-dependent Schr\"{o}dinger equations (see e.g. \cite{bao2012mathematical,bao2002time,besse2002order,lubich2008splitting}).
In particular, we mention two recent work that are more related to this paper.
In \cite{li2020numerical}, the quantum dynamics of some quasiperiodic systems were studied, by using plane wave methods for spacial discretization and an operator splitting method for time stepping to handle the kinetic and potential parts separately.
In \cite{chen2019high}, an operator splitting technique was adopted to decompose a four dimensioanl Wigner equation into two sub-equations with lower dimension, which are then discretized by a characteristic method for space-time variables and a plane wave method for momentum-time variables.

The rest of this paper is organized as follows. 
In Section \ref{sec:SEincommensurate}, a brief introduction of the linear time-dependent Schr\"{o}dinger equations for incommensuate systems is given. 
In Section \ref{sec:planewave}, a semidiscrete problem is derived by using the plane wave method for spacial discretization. 
In Section \ref{sec:temperoral discretization}, some time stepping schemes together with their convergence theory are provided. 
In Section \ref{sec:numerics}, some numerical experiments are presented to show the efficiency of the algorithms.
In Section \ref{sec:conclusion}, some conclusions are drawn.
For simplicity of the presentations, the standard operator splitting methods and the proofs of our theory are put in Appendices.

\section{Schr\"{o}dinger equation for incommensurate systems}
\label{sec:SEincommensurate}

We consider two $d$-dimensional ($d=1,2$) periodic systems that are stacked in parallel along the $(d+1)$th dimension. 
We neglect the $(d+1)$th dimension and the distance between the two layers for simplicity. 
The theories and algorithms developed in this paper can be generalized directly to incommensurate systems with more than two layers and the models involving the $(d+1)$-th dimension. 

A $d$-dimensional periodic system can be described by a Bravais lattice
\begin{eqnarray*}
\RL_j=\{A_jn:n\in \mathbb{Z}^d\},\quad j=1,2,
\end{eqnarray*}
where $A_j\in \mathbb{R}^{d\times d}$ is an invertible matrix. 
The unit cell for the $j$-th layer is given by
\begin{eqnarray*}
\Gamma_j=\{A_j\alpha:\alpha\in[0,1)^d\},\quad j=1,2.
\end{eqnarray*}
Each individual layer $\RL_j ~(j=1,2)$ is periodic in the sense that it is translation invariant with respect to its lattice vectors
\begin{eqnarray*}
\RL_j = \RL_j + A_j n \qquad \forall ~n\in \mathbb{Z}^d.
\end{eqnarray*}
The associated reciprocal lattice and reciprocal unit cell are given by
\begin{eqnarray*}
\RL_j^*=\{2\pi A_j^{-T}n:n\in \mathbb{Z}^d\}
\qquad{\rm and}\qquad
\Gamma_j^*=\{2\pi A_j^{-T}\alpha:\alpha\in[0,1)^d\}
\end{eqnarray*}
respectively. 

Though both layers $\RL_1$ and $\RL_2$ are periodic, as they are stacked together, the joined system $\RL_1\cup \RL_2$ may lose the translation invariance property, which gives the so-called incommensurate systems.

\begin{definition}(Incommensurateness).
	Lattices $\RL_1$ and $\RL_2$ are incommensurate if
	\begin{eqnarray*}
	\gamma+\RL_1^*\cup\RL_2^*=\RL_1^*\cup\RL_2^*
	\quad \Leftrightarrow \quad
	\gamma=  0\in \mathbb{R}^d.
	\end{eqnarray*}
	Otherwise, the lattices $\RL_1$ and $\RL_2$ are commensurate.
\end{definition}

We consider the following linear time-dependent Schr\"{o}dinger equation for an incommensurate system with two periodic components:
Find $\psi:\R^+\times\R^d\rightarrow \C$ such that
\begin{eqnarray}
\label{LSE}
\left\{
\begin{array}{rcll}
\displaystyle i\frac{\partial}{\partial t}\psi(t,x) &=&
\displaystyle  \Big(-\frac{1}{2}\Delta +v_1(x)+v_2(x)\Big)\psi(t,x),\quad  
& x \in\mathbb{R}^d,\ t> 0,
\\[1ex]
\psi(0,x) &=& \psi_0(x),  \ \quad \quad \ \quad  \quad \quad \quad \qquad\qquad\quad\  
& x \in\mathbb{R}^d,
\end{array}
\right.
\end{eqnarray}
where
$v_1\in L^2_{{\#,1}}(\Gamma_1) $ and $v_2 \in L^2_{{\#,2}}(\Gamma_2)$ are periodic with respect to the two lattices $\RL_1$ and $\RL_2$ respectively.
Note that the periodic Lebesgue space $L^2_{{\#,j}}(\Gamma_j)$ are defined by  
\begin{equation*}
L^2_{{\#,j}}(\Gamma_j)=\{v\in L_{{\rm loc}}^2(\R^d) \ | \ v \ \RL_j-{\rm periodic}\}\qquad {\rm for}\ j=1,2.    
\end{equation*}
We will assume $ \RL_1$ and $ \RL_2$ are incommensurate throughout this paper.
Due to the periodicity, the potential $v_j$ can be written by the expansion of plane waves $\big\{e^{iG_{jm}\cdot x}\big\}_{m\in\Z^d}$
\begin{equation}
\label{Vfourier}
v_j(x)=\sum_{m\in \mathbb{Z}^d}\hat{v}^{(j)}_m e^{iG_{jm}\cdot x}, \qquad \text{for}\  j=1,2
\end{equation}
with $G_{jm}=2\pi A_j^{-T}m\in  \RL_j^*~(m\in\Z^d)$ the wavevectors in the $j$-th reciprocal lattice and 
\begin{equation*}
\hat{v}^{(j)}_m=\frac{1}{|\Gamma_{j}|}\int_{\Gamma_j}v_{j}(x)e^{-iG_{jm}\cdot x}\mathrm{d}x.
\end{equation*}
This type of incommensurate problems are commonly used to study the ultracold atoms \cite{albert2010localization,roati2008anderson} and the photonic crystals \cite{lahini2009observation,wang2020localization}.

An important feature of the solutions of \eqref{LSE} is the conservation of mass and energy \cite{gottfried2013quantum}.
Let
\begin{align}
\label{massN}
N(\psi(\cdot,t)) &:= \lim\limits_{R\rightarrow\infty}\frac{1}{B_R}\int_{B_R}|\psi(t,x)|^2\mathrm{d} x, 
\\[1ex]
\label{energyE}
E(\psi(\cdot,t)) &:= \lim\limits_{R\rightarrow\infty}\frac{1}{B_R}\int_{B_R}
\Big(\frac{1}{2}|\nabla\psi(t,x)|^2 + v_1(x)|\psi(t,x)|^2 + v_2(x)|\psi(t,x)|^2\Big) 
\mathrm{d} x 
\end{align}
denote the {\it averaged} mass and energy of the wave function $\psi(t,\cdot)$ respectively, where $B_R\subset \R^d$ is the ball centred at origin with radii $R$.
Then these quantities are conserved during the evolution, more precisely, we have
\begin{eqnarray*}
N(\psi(\cdot,t)) \equiv N(\psi_0(\cdot,t))
\qquad{\rm and}\qquad
E(\psi(\cdot,t)) \equiv E(\psi_0(\cdot,t)) .
\end{eqnarray*}

\section{Semidiscrete problem by plane wave discretization}
\label{sec:planewave}

Following the ideas of \cite{jiang2014numerical,zhou2019plane}, we use the plane wave methods for spatial discretization of the Schr\"{o}dinger equation \eqref{LSE}, to obtain a semidiscrete ordinary differential equation system.

Let $\Ec>0$ be the energy cutoff of the plane wave discretization.
Let $\Index{1}$ and $\Index{2}$ be the set of wavevectors for the lattices $\RL_1$ and $\RL_2$ respectively
\begin{eqnarray*}
\Index{1}=\{m \in \Z^d:~ |G_{1m}|^2\leq2\Ec\}
\qquad {\rm and} \qquad 
\Index{2}=\{n \in \Z^d:~ |G_{2n}|^2\leq2\Ec\}
\end{eqnarray*}
with their cardinalities denoted by $|\Index{1}|$ and $|\Index{2}|$.

We use the following plane waves as basis functions for spatial discretization
\begin{equation*}
\label{basis_function}
\bigg\{e^{i(G_{1m}+G_{2n})\cdot x}\bigg\}_{(m,n)\in \mathcal{I}_{\Ec}}
\qquad{\rm with}\quad
\Index{} := \Index{1}\times\Index{2} = \big\{(m,n):~ m\in \Index{1},\ n\in \Index{2}\big\} .
\end{equation*}
We immediately have the following orthonormal condition:
\begin{eqnarray}
\label{ortho}
\intRd e^{-i(G_{1m}+G_{2n})x} e^{i(G_{1m'}+G_{2n'})x} \dd x = \delta_{mm'}\delta_{nn'} \qquad\forall~(m,n),(m',n')\in\Index{} ,
\end{eqnarray}
where the ``averaged" spacial integral is defined by 
\begin{eqnarray*}
\intRd:=\lim_{R\rightarrow\infty} \frac{1}{|B_R|}\int_{B_R} 
\end{eqnarray*}
with $B_R\subset\R^d$ the ball centred at origin with radii $R$.
The number of basis functions is determined by the energy cutoff $\Ec$, 
more precisely, $|\Index{}|=|\Index{1}|\cdot|\Index{2}| \sim \mathcal{O}\big(\Ec^d\big)$.
A larger energy cutoff will lead to more accurate results, but at the price of increasing the computational cost significantly.
  
With the the plane wave basis set, we can approximate the solution $\psi(t,x)$ of \eqref{LSE} by
\begin{equation}
\label{wavefunctionapprox}
\psi_{\Ec}(t,x)=\sum_{(m,n)\in \mathcal{I}_{\Ec} }\phi_{mn}(t)e^{i(G_{1m}+G_{2n})\cdot x},
\end{equation}
where the coefficients $\phi_{mn}(t)~\big((m,n)\in\Index{}\big)$ are time-dependent functions.
Then we can derive the semidiscrete equations for $\phi_{mn}(t)$ from \eqref{LSE} by using standard Galerkin projection and the orthonormal condition \eqref{ortho}
\begin{equation}
\label{ode}
\frac{\mathrm{d}\phi_{mn}(t)}{\mathrm{d}t} 
=  -i\ \Bigg(\frac{1}{2}|G_{1m}+G_{2n}|^2\phi_{mn}(t)+\sum_{m'\in \mathbb{Z}^d}\hat{v}^{(1)}_{m'}\phi_{{m-m'},n}(t)+\sum_{n'\in \mathbb{Z}^d}\hat{v}^{(2)}_{n'}\phi_{m,{n-n'}}(t)\Bigg) ,
\end{equation}
where the initial state $\{\phi_{mn}(0)\}_{(m,n)\in\Index{}}$ can be chosen by some interpolation/projection of $\psi_0(x)$ (in \eqref{LSE}) with the plane wave basis functions.

Denote by $\vecn{}(t)=(\phi_{mn}(t))^T_{(m,n)\in \Index{}}$, the ordinary differential equations system \eqref{ode} can be rewritten in the matrix form
\begin{equation}
\label{ODE}
\displaystyle \frac{\mathrm{d}\vecn {}(t)}{\mathrm{d}t}=-i H\vecn {}(t),
\end{equation}
where $H \in \C ^{| \Index{}|\times|\Index{}|}$ is a Hermitian matrix
\begin{align}
\label{H}
& H = D+V^1+V^2
\qquad\qquad{\rm with}
\\[1ex]
\label{DV1V2}
& D_{mn,m'n'}=\frac{1}{2}|G_{1m}+G_{2n}|^2\delta_{mm'}\delta_{nn'},\quad
V^1_{mn,m'n'}=\vm{(m-m')}\delta_{nn'} \quad
{\rm and}\quad
V^2_{mn,m'n'}=\vn{(n-n')}\delta_{mm'} 
\end{align}
for $(m,n),(m',n')\in\Index{}$.
In the following sections, we will denote the initial state $\vecn {}(0)$ by $\vecn {0}$ to illustrate the time stepping schemes.

\begin{figure}[H]
\centering
\subfigure[$V^1+V^2$]{
	\label{v1v2structure}
	\includegraphics[height=3.0cm,width=3.8cm]{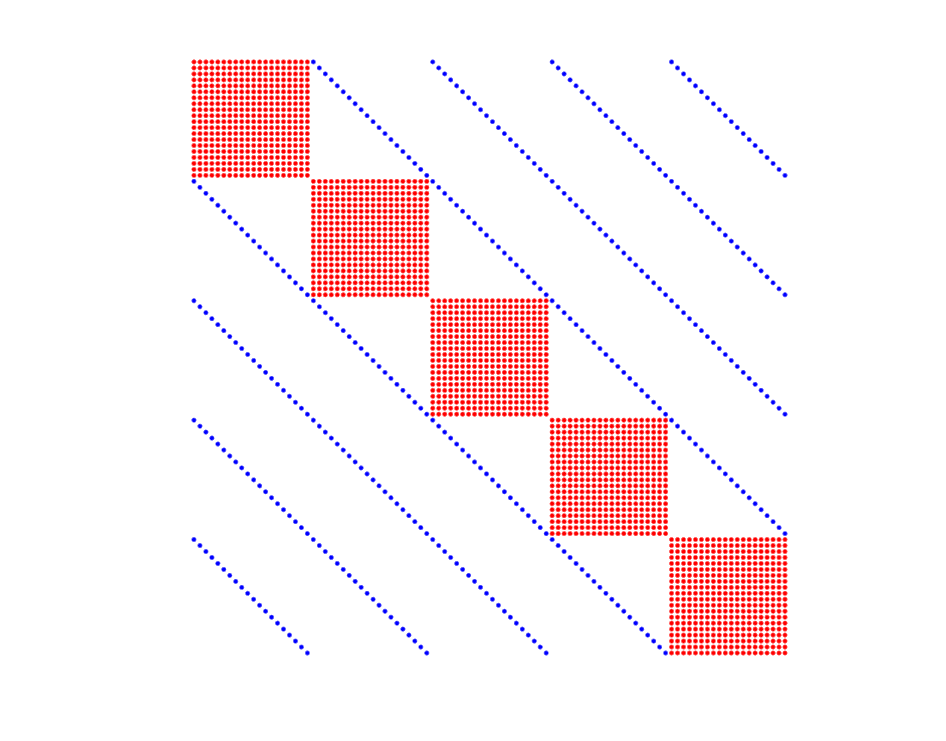}}
\subfigure[$V^1$]{
	\label{v1structure}
	\includegraphics[height=3.0cm,width=3.8cm]{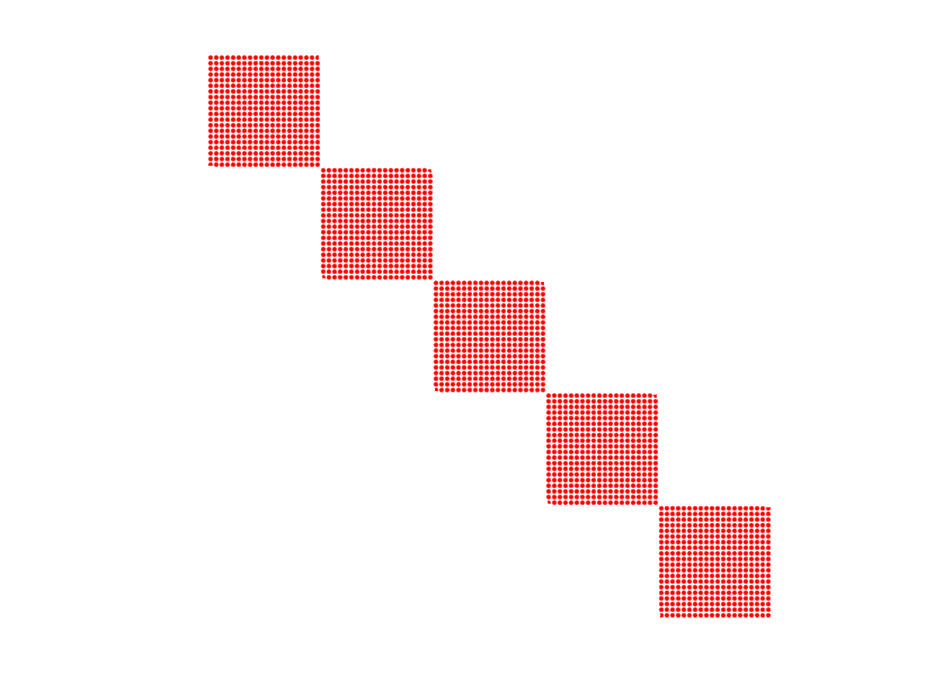}}
\subfigure[$V^2$]{
	\label{v2structure}
	\includegraphics[height=3.0cm,width=3.8cm]{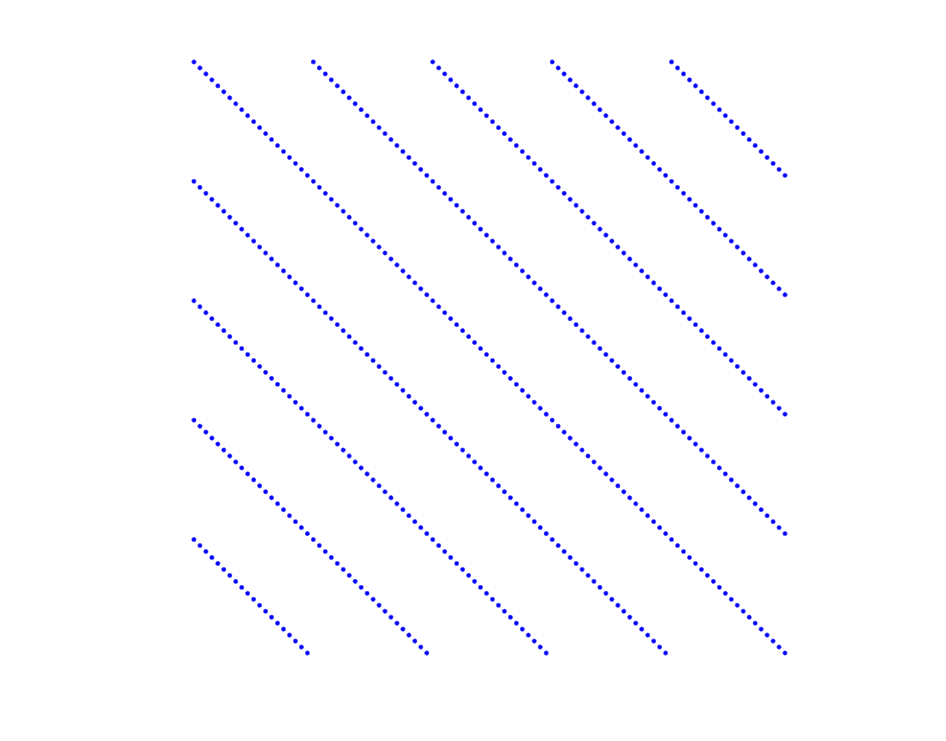}}
\subfigure[Permuted $V^2$]{
	\label{blockv2structure}
	\includegraphics[height=3.0cm,width=3.8cm]{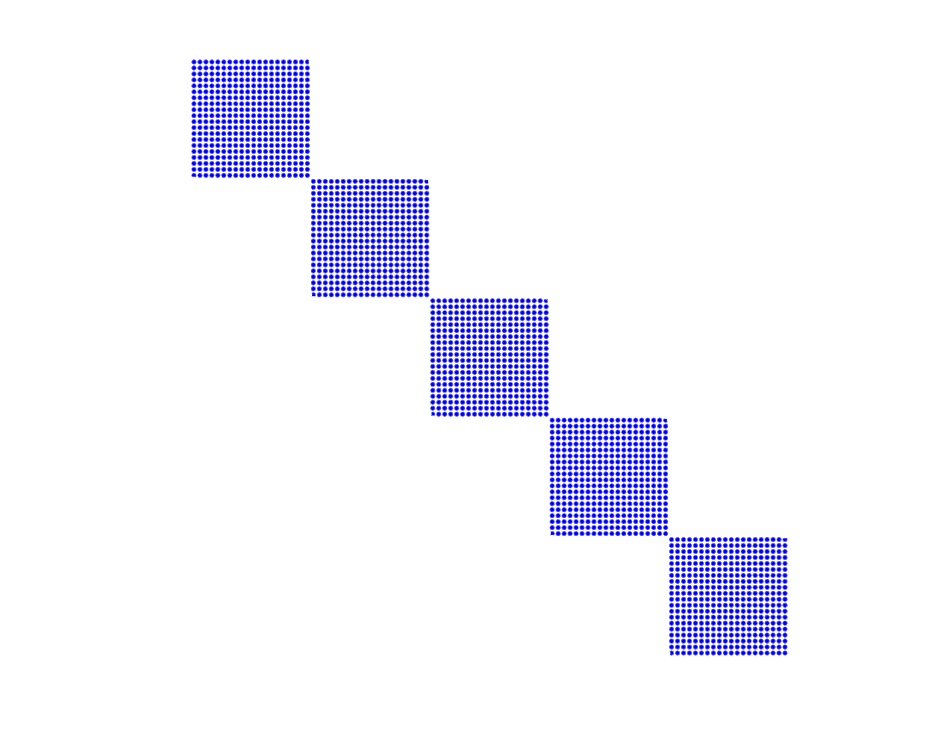}}	
\caption{Structures of the matrices $V^1$ and $V^2$.}
\label{matrixstructures}
\end{figure}

The dimension of the matrix $H$ is huge in this plane wave representation, since it essentially lift the problem into a higher dimensional system (see \cite{zhou2019plane}).
To be more specific, the matrix $D$ related to the Laplace operator is diagonal, yet the matrix $V$ related to the incommensurate potentials is what makes the computations complicated.
We observe that the matrices $V^1$ and $V^2$ can be written in the following form of Kronecker products
\begin{eqnarray*}
\label{V1V2}
V^1=I_1\otimes \Vc^{1} \qquad \text{and} \qquad  V^2=\Vc^{2}\otimes I_2 ,
\end{eqnarray*}
where
$I_1\in\R^{|\Index{1}|\times|\Index{1}|}$ and $I_2\in \R^{|\Index{2}|\times|\Index{2}|}$ denote the identity matrices, 
$\Vc^{1}\in  \C^{|\Index{1}|\times|\Index{1}|}$ and $\Vc^{2}  \in  \C^{|\Index{2}|\times|\Index{2}|}$ are plane wave representations of the two periodic potentials respectively
\begin{equation*}
\label{V1nV2m}
{\Vc^{1}}_{mm'}=\vm{(m-m')}\qquad \text{and} \qquad {\Vc^{2}}_{nn'}=\vn{(n-n')}
\end{equation*}
with $\vj~(j=1,2)$ defined in \eqref{Vfourier}.
Based on the above representation of Kronecker products, we see that $V$ is essentially a sum of two ``block diagonal" matrices, which are block diagonal for $V^1$ and block diagonal after some permutation for $V^2$.
We show in Figure \ref{matrixstructures} a schematic plot of the matrix structures. 
Note that this block diagonal matrix structure (after splitting the incommensurate layers into periodic ones and permutations) can be directly generalized to systems with more than two layers.

The above observations motivate us to design efficient time stepping schemes for the semidiscrete problem \eqref{ode} to reduce the computational cost of the high dimensional problem.
The idea is to decompose the problem into sub-problems, each corresponds to a periodic potential, such that the block diagonal matrix structure can be exploited in the algorithm.
It is a ``layer-splitting" method, in the sense that for each sub-problem, one of the periodic layers are dealt with rather than the whole incommensurate systems.

\section{Time stepping schemes based on layer-splitting}
\label{sec:temperoral discretization}

With a given energy cutoff $\Ec$ for plane wave discretizations, a standard time stepping scheme for solving \eqref{ode} by directly using matrix $H$ scales as $\Oc\big(|\Index{}|^2\big)=\Oc\big(\Ec^{2d}\big)$ for each step, due to the sizes of $H$.
Note that for systems with $L(\geq 2)$ layers, the computational cost scales like $\Oc\big(\Ec^{dL}\big)$ within the plane wave framework.
In this section, we will present some numerical schemes based on the idea of layer-splitting to reduce the computational cost, such that it does not increases exponentially fast with respect to the number of layers.

Let $0=t_0<t_1<\dots < t_K=T$ be a uniform discretization of the considered time interval $[0,T]$, and $\tau=t_k-t_{k-1}~ (1\leq k \leq K)$ be the time step. 
Denote by $\phi_{mn}^k$ the approximation of $\phi_{mn}(t_k)$ and $\vecn {k}$ the vector collecting all components $\phi_{mn}^k$.
In our analysis, we will denote by $C_T$ a general constant depending on $T$, but independent of the discretization parameter $\tau$.

\subsection{Semi-implicit method}
\label{sec:semiimplicit}

We first propose a semi-implicit form for \eqref{ODE}, by treating the kinetic part $D$ implicitly and the potential parts $V^j~(j=1,2)$ explicitly.
More precisely, we have the following fully discrete form for \eqref{ODE}
\begin{eqnarray}
\label{semi-implicitV1V2}
\displaystyle\frac{\vecn {k+1}-\vecn {k}}{\tau}=-i \Big( D\ \vecn {k+1}+(I_1\otimes \Vc^{1})\vecn{k}+(\Vc^{2}\otimes I_2)\vecn{k}\Big)
\end{eqnarray}
for $k=0,\cdots,K-1$, in which we have replaced $V^j~(j=1,2)$ by their Kronecker product form.

We can rewrite \eqref{semi-implicitV1V2} by the time stepping scheme
\begin{eqnarray*}
\label{semi-implic}
\vecn{k+1} = (I+ i\tau D)^{-1}\big(\vecn{k} - i\tau(I_1\otimes \Vc^{1})\vecn{k} - i\tau(\Vc^{2}\otimes I_2)\vecn{k}\big) .
\end{eqnarray*}
Here, the inverse of the matrix does not bring in any more difficulty since $D$ is a diagonal matrix.
Since the potential parts are dealt with explicitly, the matrix products with $I_1\otimes \Vc^{1}$ and $\Vc^{2}\otimes I_2$ can be done separately and the Kronecker product form (i.e. the block diagonal structure of the matrix shown in Section \ref{sec:planewave}) can be exploited.
Then we can write down the detailed algorithm in the following.

\begin{algorithm}
\caption{: Semi-implicit method}
\label{alg:semiimplicit}
\begin{algorithmic}
\State Given a time step $\tau>0$
\For { $k=0: K-1$ }  
\State 1:  
Compute $\Psi_{1} := \big(\psi_{1,mn}\big)_{(m,n)\in\Index{}}$ with
$\big(\psi_{1,mn}\big)_{m\in\Index1} = i\tau \Vc^1(\phi_{mn}^k)_{m\in\Index1}$ for $n\in\Index2$.
\State  2: 
Compute $\Psi_{2} := \big(\psi_{2,mn}\big)_{(m,n)\in\Index{}}$ with
$\big(\psi_{2,mn}\big)_{n\in\Index2} = i\tau \Vc^2(\phi_{mn}^k)_{n\in\Index2}$ for $m\in\Index1$.
\State 3: 
Compute $\vecn{k+1}=(I+i\tau  D)^{-1}\big(\vecn{k}-\Psi_{1}-\Psi_{2}\big)$.
\EndFor
\end{algorithmic}
\end{algorithm}

We then look into the computational cost of Algorithm \ref{alg:semiimplicit} for each time step.
The explicit part will require multiplications with matrix $\Vc^1$ for $|\Index2|$ times and multiplications with matrix $\Vc^2$ for $|\Index1|$ times, whose cost scales like $\Oc(|\Index1|^2\cdot|\Index2|+|\Index2|^2\cdot|\Index1|) = \Oc(\Ec^{3d/2})$. 
The cost for implicit part scales like $\Oc(|\Index{}|)=\Oc(\Ec^d)$ since $D$ is diagonal.
Therefore, the total cost for Algorithm \ref{alg:semiimplicit} scales as $\Oc(\Ec^{3d/2})$, which reduces the cost significantly, compared with that of the standard scheme (i.e. $\Oc(\Ec^{2d})$) without splitting $V^1$ and $V^2$.
We mention that the cost can be further reduced if the fast Fourier transform is applied to the multiplication with $\Vc^1$ and $\Vc^2$, see Section \ref{sec:tsF}.

We see that this algorithm is naturally compatible with parallel computing since the matrix-vector multiplication can be implemented on many nodes for different blocks.

The following theorem gives the convergence rate of this semi-implicit method.
The key is that the kinetic part $D$ has a larger spectral radius compared with the smooth potential parts, and therefore treating $D$ implicitly can ensure the time stepping scheme to be stable.
The proof is relatively standard and is given in Appendix \ref{sec:appendix proofs} for completeness of this paper.

\begin{theorem}\label{error_semi}
	Let $\vecn{}(t)$ be the solution of \eqref{ODE} and $\vecn k$ be obtained by Algorithm \ref{alg:semiimplicit}.
	Then there exist positive constants $\tau_0$ and $C_T$ such that
	\begin{equation}\label{errorbound_semi}
		||\vecn{}(t_k)-\vecn k||_{\ell_2}\leq C_T\tau 
	\end{equation}	
	for any time step $0<\tau<\tau_0$ and $t_k=k\tau\in [0,T]$.
\end{theorem}

Although the semi-implicit method is efficient and easy to implement, it is a first order method and hence requires small time step to obtain high accuracy results.
Moreover, we will see in our numerical experiments (in Section \ref{sec:numerics}) that it does not conserve the mass of the wavefunctions.
These motivate us to develop higher order and mass conserved time stepping schemes.

\subsection{Operator splitting methods}
\label{sec:tsbe}

Since the matrix $H$ in \eqref{H} can be expressed as
\begin{equation*}\label{Hsplit}
	H=\left(\frac{1}{2}D+I_1\otimes \Vc^{1}\right)+\left(\frac{1}{2}D+\Vc^{2}\otimes I_2\right),
\end{equation*}
the semidiscrete problem \eqref{ODE} can be split into two sub-problems by the so-called operator splitting methods \cite{glowinski2017splitting,marchuk1968some,strang1968construction}, each part corresponds to one of the periodic layers.
We briefly review the standard operator splitting methods in Appendix \ref{sec:appendix splitting method} for completeness.

\subsubsection{Backward Euler and  Crank-Nicolson based operator splitting methods}
\label{sec:osplitting}

We first apply the Lie-Trotter splitting scheme (see Appendix \ref{sec:appendix splitting method}), together with the backward Euler method to \eqref{ODE}, to obtain the fully discrete form
\begin{equation*}
\label{tsbE}
\left\{
\begin{array}{l}
\displaystyle \frac{\vecn{k+1/2}-\vecn{k}}{\tau}=-i \Big(\frac{1}{2}D+(I_1\otimes \Vc^{1})\Big)\vecn{k+1/2},
\\[2ex]
\displaystyle \frac{\vecn{k+1}-\vecn{k+1/2}}{\tau}=-i \Big(\frac{1}{2}D+(\Vc^{2}\otimes I_2)\Big)\vecn{k+1}
\end{array}
\right.		
\end{equation*}
for $k=0,\dots,K-1$.
This can be written by the following time stepping schemes
\begin{align}
\label{tsbEtimestepping1}
& \vecn{k+1/2}=\Big(I+i\tau\big(\frac{1}{2}D+(I_1\otimes \Vc^{1})\big)\Big)^{-1}\vecn{k},
\\[1ex]
\label{tsbEtimestepping2}
& \vecn{k+1}=\Big(I+i\tau\big(\frac{1}{2}D+(\Vc^{2}\otimes I_2)\big)\Big)^{-1}\vecn{k+1/2}.
\end{align}

By exploiting the Kronecker product structure of the matrix, the multiplication with the inverse of the matrices in \eqref{tsbEtimestepping1} and \eqref{tsbEtimestepping2} can be performed by calculating the inverse of each block separately.
The detailed algorithm is given as follows, in which we use the notations $D_n:=\big(D_{mn,mn}\big)_{m \in \Index1} \in\R^{|\Index1|\times|\Index1|}$ and $D_m:=\big(D_{mn,mn}\big)_{n \in \Index2} \in\R^{|\Index2|\times|\Index2|}$.

\begin{algorithm}[H]
\caption{: Backward Euler based operator splitting method}
\label{alg:tsbE}
\begin{algorithmic}
\State Given a time step $\tau>0$
\For { $k=0: K-1$ }  
\State 1:  
Compute $\vecn{k+1/2} = \big(\phi_{mn}^{k+1/2}\big)_{(m,n)\in\Index{}}$ with
\vskip -0.3cm
\begin{eqnarray*}
\big(\phi_{mn}^{k+1/2}\big)_{m\in\Index1} = \Big(I_1+i\tau\big( D_n/2 + \Vc^1\big)\Big)^{-1} (\phi_{mn}^k)_{m\in\Index1}\quad {\rm for} \ n\in \Index2.
\end{eqnarray*}
\State  2: 
Compute $\vecn{k+1} = \big(\phi_{mn}^{k+1}\big)_{(m,n)\in\Index{}}$ with
\vskip -0.3cm
\begin{eqnarray*}
\big(\phi_{mn}^{k+1}\big)_{n\in\Index2} = \Big(I_2+i\tau\big( D_m/2 + \Vc^2\big) \Big)^{-1} (\phi_{mn}^{k+1/2})_{n\in\Index2}\quad {\rm for} \ m\in \Index1.
\end{eqnarray*}
\EndFor
\end{algorithmic}
\end{algorithm}

In each time step of Algorithm \ref{alg:tsbE}, step 1 requires solving $|\Index2|$ linear systems of order $|\Index1|$ and step 2 requires solving $|\Index1|$ linear systems of order $\Oc(|\Index2|)$.
Therefore, the total computational cost for Algorithm \ref{alg:tsbE} is $\Oc(|\Index1|^2\cdot|\Index2|) + \Oc(|\Index2|^2\cdot|\Index1|) = \Oc(\Ec^{3d/2})$. 
More importantly, these linear systems can be solved in parallel. 
The following theorem gives the convergence rate of the backward Euler based operator splitting method, whose proof is given in Appendix \ref{sec:appendix proofs}.

\begin{theorem}
\label{error_splittingbE}
	Let $\vecn{}(t_k)$ be the solution of \eqref{ODE} , and $\vecn k$ be obtained by Algorithm \ref{alg:tsbE}. Then there exist positive constants $\tau_0$ and $C_T$ such that
	\begin{equation}\label{errorbound_splittingbE}
		||\vecn{}(t_k)-\vecn k||_{\ell_2}\leq C_T\tau
	\end{equation}	
    for any time step $0<\tau<\tau_0$ and $t_k=k\tau\in [0,T]$.
\end{theorem}

We see from this theorem that the backward Euler based operator splitting method is a first order method, the same as the semi-implicit method. 
We will then construct some second order method to improve the accuracy. 
To do this, we combine the Strang splitting scheme (see Appendix \ref{sec:appendix splitting method}) and the  Crank-Nicolson method to obtain the following fully discrete form for \eqref{ODE}

\begin{equation}
\label{tsCN}
\left\{
\begin{array}{l}
\displaystyle \frac{\vecn{k+1/3}-\vecn{k}}{\tau/2}=-i \Big(\frac{1}{2}D+(I_1\otimes \Vc^{1})\Big)\displaystyle\frac{\vecn{k+1/3}+\vecn{k}}{2},
\\[2ex]
\displaystyle \frac{\vecn{k+2/3}-\vecn{k+1/3}}{\tau}=-i \Big(\frac{1}{2}D+(\Vc^{2}\otimes I_2)\Big)\displaystyle\frac{\vecn{k+2/3}+\vecn{k+1/3}}{2},
\\[2ex]
\displaystyle \frac{\vecn{k+1}-\vecn{k+2/3}}{\tau/2}=-i \Big(\frac{1}{2}D+(I_1\otimes \Vc^{1})\Big)\displaystyle\frac{\vecn{k+1}+\vecn{k+2/3}}{2}
\end{array}
\right.		
\end{equation}
for $k=0,\dots,K-1$. 
This can be rewritten by the time stepping scheme
\begin{align*}
&\vecn{k+1/3}=\Big(4I+i\tau\big(D/2+I_1\otimes \Vc^{1}\big)\Big)^{-1}\Big(4I-i\tau\big(D/2+I_1\otimes \Vc^{1}\big)\Big)\vecn{k},  \\
&\vecn{k+2/3}=\Big(2I+i\tau\big(D/2+\Vc^{2}\otimes I_2\big)\Big)^{-1}\Big(2I-i\tau\big(D/2+\Vc^{2}\otimes I_2\big)\Big)\vecn{k+1/3},    \\
&\vecn{k+1}=\Big(4I+i\tau\big(D/2+I_1\otimes \Vc^{1}\big)\Big)^{-1}\Big(4I-i\tau\big(D/2+I_1\otimes \Vc^{1}\big)\Big)\vecn{k+2/3}.
\end{align*}

We will skip the details of the algorithm to explain how to implement the above Crank-Nicolson based operator splitting method, which is similar to that of the backward Euler method. 
The following theorem gives the convergence rate of the splitting scheme \eqref{tsCN}, whose proof is given in Appendix \ref{sec:appendix proofs}.

\begin{theorem}
\label{error_splittingCN}
	Let $\vecn{}(t_k)$ be the solution of \eqref{ODE} , and $\vecn k$ be obtained by \eqref{tsCN}. Then there exist positive constants $\tau_0$ and $C_T$ such that
	\begin{equation}\label{errorbound_splittingCN}
		||\vecn{}(t_k)-\vecn k||_{\ell_2}\leq C_T\tau^2
	\end{equation}	
	for any time step $0<\tau<\tau_0$ and $t_k=k\tau\in [0,T]$.
\end{theorem}

Moreover, we have that Crank-Nicolson based operator splitting time stepping scheme can conserve the mass during the evolution, which is shown in the next theorem.
The proof is referred to Appendix \ref{sec:appendix proofs}.

\begin{theorem}
\label{massconservationtsCN}
    Let $\Ec>0$, $\vecn{k} = \big(\phi^k_{mn}\big)_{(m,n)\in\Index{}} \in\R^{|\Index{}|}$ be obtained by the fully discrete scheme \eqref{tsCN},
    and $\displaystyle \psi_{\Ec}^k(x) = \sum_{(m,n)\in \mathcal{I}_{\Ec} }\phi^k_{mn}e^{i(G_{1m}+G_{2n})\cdot x}$.
    Then
	\begin{equation}
	\label{mass conservation tsCN}
	N\big(\psi_{\Ec}^k\big) = N\big(\psi_{\Ec}^0\big) \qquad{\rm for}~ 0\leq k\leq K ,
	\end{equation}
	where $N(\cdot)$ is defined in \eqref{massN}.
\end{theorem}

\subsubsection{Fourier transform based operator splitting method}
\label{sec:tsF}

We observe that the Hamiltonian in the Schr\"{o}dinger equation \eqref{LSE} consists of the kinetic part $-\frac{1}{2}\Delta$ and the potential part $v_1+v_2$, which are (diagonal and hence) easy to implement in the momentum space and real space, respectively.
Based on this observation, we can construct operator splitting methods that split the kinetic and potential parts in the time stepping scheme, so that the computational cost can be further reduced.

By using the Lie-Trotter splitting scheme \eqref{splitA}, \eqref{ODE} can be split into the following three sub-problems
\begin{equation}
\label{tsF}
\left\{
\begin{array}{l}
\vecn{k+1/3}=e^{-i\tau D}\vecn{k},
\\[2ex]
\vecn{k+2/3}=e^{-i\tau V^1}\vecn{k+1/3},
\\[2ex]
\vecn{k+1}=e^{-i\tau V^2}\vecn{k+2/3}.
\end{array}
\right.		
\end{equation}
We can then apply the fast Fourier transform (FFT) and inverse fast Fourier transform (iFFT) algorithms to switch between the real and momentum spaces, so that all exponents in \eqref{tsF} can be evaluated exactly and efficiently.
Let $\Fc:\R^n\rightarrow\R^n$ and $\Fc^{-1}:\R^n\rightarrow\R^n$ ($n\in\Z^+$) denote the FFT and iFFT transforms respectively.
We present the details of our Fourier transform based operator splitting method in the following algorithm.

\begin{algorithm}[H]
	\caption{: Fourier transform based operator splitting method}
	\label{alg:tsF}
	\begin{algorithmic}
	\State Given a time step $\tau>0$
	\vskip 0.2cm	  
	 \State 
	 Perform iFFT: ~
	 $(v^{(1)}_m)_{m\in \Index1}=\Fc^{-1}\big((\vm{m})_{m\in \Index1}\big)$
	 and ($v^{(2)}_n)_{n\in \Index2}=\Fc^{-1}\big((\vn{n})_{n\in \Index2}\big)$
    \vskip 0.2cm
		\For { $k=0: K-1$ }
		\State 1:  
		Compute ~ $\vecn{k+1/3}:=(\phi_{mn}^{k+1/3})_{(m,n)\in \Index{}}$ with
		$$\phi_{mn}^{k+\frac{1}{3}}=\exp(-i\tau |G_{1m}+G_{2n}|^2/2)\phi_{mn}^{k}\quad {\rm for}\ (m,n)\in \Index{}.$$
		\State 2: 
		Perform iFFT: ~ $(\psi_{mn}^{k+1/3})_{m\in \Index1}:=\Fc^{-1}\big((\phi_{mn}^{k+1/3})_{m\in \Index1}\big)\quad {\rm for}\ n\in \Index2.$
		\State 3: 
		Compute ~ $\Psi^{k+2/3}:=(\psi_{mn}^{k+2/3})_{(m,n)\in \Index{}}$ with
		 $$\psi_{mn}^{k+2/3}=\exp(-i\tau v^{(1)}_m)\psi_{mn}^{k+1/3}\quad {\rm for}\ (m,n)\in \Index{}. $$
		\State 4: 
		  Compute ~ $\Psi^{k+1}:=(\psi_{mn}^{k+1})_{(m,n)\in \Index{}}$ with
		 $$\psi_{mn}^{k+1}=\exp(-i\tau v^{(2)}_n)\psi_{mn}^{k+2/3}\quad {\rm for}\ (m,n)\in \Index{}. $$
        \State 5:
        Perform FFT: ~ $(\phi_{mn}^{k+1})_{m\in \Index1}=\Fc\big((\psi_{mn}^{k+1})_{m\in \Index1}\big)\quad {\rm for}\ n\in \Index2.$
        \EndFor
	\end{algorithmic}
\end{algorithm}

In Algorithm \ref{alg:tsF}, the computational costs for both kinetic and potential parts scale like $\Oc(|\Index{}|)=\Oc(\Ec^d)$,
the costs for FFT and iFFT scale like $\Oc\big(|\Index{}| (\ln(|\Index1|)+\ln(|\Index1|)\big) = \Oc(\Ec^{d}\ln(\Ec))$.
Therefore the total cost for Algorithm \ref{alg:tsF} is $\Oc(\Ec^d\ln(\Ec))$ for each time step, which is better than our previous schemes.
Particularly, parallel computing can be adopted naturally in this algorithm. 

The following theorems show the convergence and mass conservation of the Fourier transform based operator splitting method, whose proof is given in Appendix \ref{sec:appendix proofs}.

\begin{theorem}\label{error_splittingtsFourier}
	Let $\vecn{}(t_k)$ be the solution of \eqref{ODE} , and $\vecn k$ be solution obtained by Algorithm \ref{alg:tsF}. Then there exist positive constants $\tau_0$ and $C_T$ such that
	\begin{equation}\label{errorbound_splittingtsFourier}
		||\vecn{}(t_k)-\vecn k||_{\ell_2}\leq C_T\tau
	\end{equation}	
	for any time step $0<\tau<\tau_0$ and $t_k=k\tau\in [0,T]$.
\end{theorem}

\begin{theorem}
\label{massconservationtsFourier}
    Let $\Ec>0$, $\vecn{k} = \big(\phi^k_{mn}\big)_{(m,n)\in\Index{}} \in\R^{|\Index{}|}$ be obtained by Algorithm \ref{alg:tsF},
    and $\displaystyle \psi_{\Ec}^k(x) = \sum_{(m,n)\in \mathcal{I}_{\Ec} }\phi^k_{mn}e^{i(G_{1m}+G_{2n})\cdot x}$.
    Then
	\begin{equation}
	\label{masstsF}
	N\big(\psi_{\Ec}^k\big) = N\big(\psi_{\Ec}^0\big) \qquad{\rm for}~ 0\leq k\leq K .
	\end{equation}
\end{theorem}

The Fourier transform based operator splitting method is efficient and easy to implement, in particular, no time discretization error is introduced for the sub-problems in \eqref{tsF} and the mass of wavefunction is conserved in the evolution.

We can further extend the algorithm to higher order time stepping schemes by exploiting the high order operator splitting methods (see e.g. \cite{hairer2006geometric}).
We will not discuss the algorithm details here, but only present some numerical experiments with the higher order time stepping schemes in Section \ref{sec:numerics}. 

\subsection{Random batch method}
\label{sec:random}

The random batch methods \cite{golse2019random,jin2020random}, are stochastic algorithms that have been widely used for simulations of large systems.
Here we propose a random batch method to solve \eqref{ODE}. 
For each time step, we randomly pick small subsets instead of the whole sets $\Index1$ and $\Index2$, and consider only the updates in each subsets. 
The detailed algorithm is given in the following.

\begin{algorithm}[H]
	\caption{: Random batch method ($p$ block(s))}
	\begin{algorithmic}
		\State 
		Given a time step $\tau>0$ and a positive constant $p<<\Ec^{d/2}$
		\For { $k=0: K-1$ }  
		\State 
		Pick randomly $P_1=\{m_1,m_2,\dots, m_p\} \subset \Index{1}$ and $P_2=\{n_1,n_2,\dots, n_p\} \subset \Index{2}$.
		\State 
		Let $\Psi_1=0 \in\R^{|\Index{}|}$ and $\Psi_2=0  \in\R^{|\Index{}|}$.
		\State 1: 
		Compute $\Psi_1:=(\psi_{1,mn})_{(m,n)\in\Index{}}$ with
        $(\psi_{1,mn})_{m\in \Index1}=\frac{|\Index2|}{p}\cdot i\tau \Vc^1(\phi_{mn}^k)_{m\in \Index1}$ for $n\in P_2$.
		\State  2: 
		Compute
		$\Psi_2:=(\psi_{2,mn})_{(m,n)\in\Index{}}$ with
        $(\psi_{2,mn})_{n\in \Index2}=\frac{|\Index1|}{p}\cdot i\tau \Vc^2(\phi_{mn}^k)_{n\in \Index2}$ for $m\in P_1$.
		\State 3: 
		Compute
		$\vecn{k+1}=(I+ i\tau  D)^{-1}(\vecn{k}-\Psi_1-\Psi_2)$.
		\EndFor
		
	\end{algorithmic}
\end{algorithm}

We mention that the sizes of sets $P_1$ and $P_2$ do not have to be equal in the above algorithm.
Though the random batch method is much cheaper than the algorithms without stochastic techniques,
we observe in our numerical experiments (in Section \ref{sec:numerics}) that their approximations possess the same convergence rates.
The theoretical analysis of this stochastic algorithm deserves to be investigated in our other work. 

\section{Numerical experiments}
\label{sec:numerics}

In this section, we will perform some numerical experiments on the time-dependent linear Schr\"{o}dinger equations of some incommensurate systems.
In each example, we will fix the energy cutoff $\Ec$ and take the following wave function as the initial state
\begin{equation*}\label{initial approximated wavefunction}
\psi_{0,\Ec}(x) = \sum_{(m,n)\in \Index{} }e^{-\gamma|G_{1m}+G_{2n}|^2}e^{i(G_{1m}+G_{2n})\cdot x},
\end{equation*} 
from which the initial $\vecn {}_0$ in \eqref{ODE} is automatically determined.
To compute the numerical error, the results obtained with a very small time step, say $\tau=0.001$, are used as the exact solutions.

{\bf Example 1.}   
(1D incommensurate systems)
Consider the following one dimensional Schr\"{o}dinger equation:
\begin{equation*}
	\displaystyle i\frac{\partial}{\partial t}\psi(t,x) =\Big(-\frac{1}{2}\Delta+s_1\cos(\beta x)^2+s_2\cos( x)^2\Big)\psi(t,x)
\end{equation*}
with $s_1=1,\ s_2=2$ and $\beta=\frac{\sqrt{5}-1}{2}$.
We take $\Ec=500$ and $\gamma=100$ in this example.

\begin{figure}[htb!]
	\centering  
	\subfigure{
		\includegraphics[height=6.0cm]{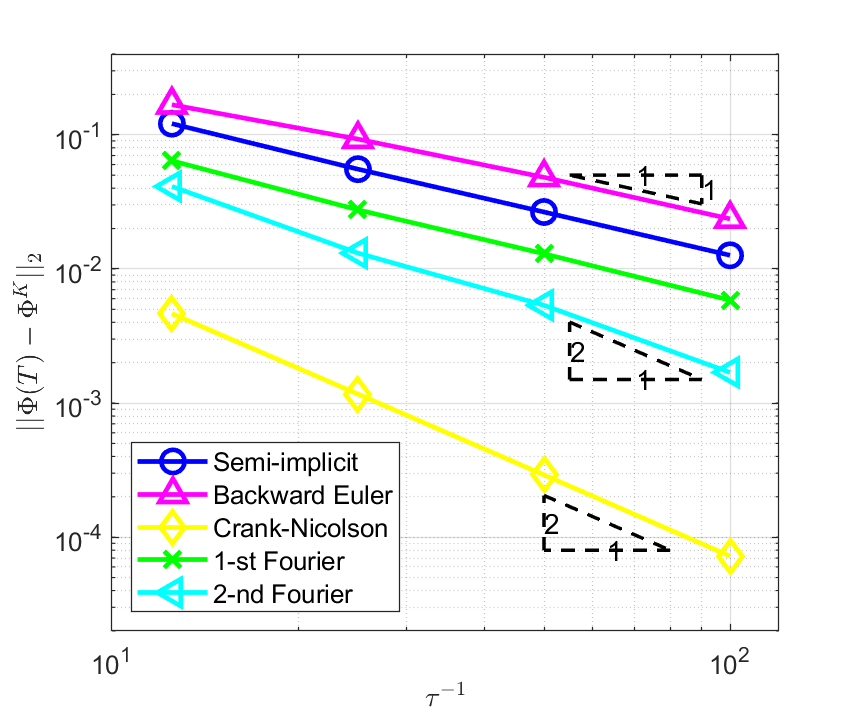}}
	\subfigure{
		\includegraphics[height=6.0cm]{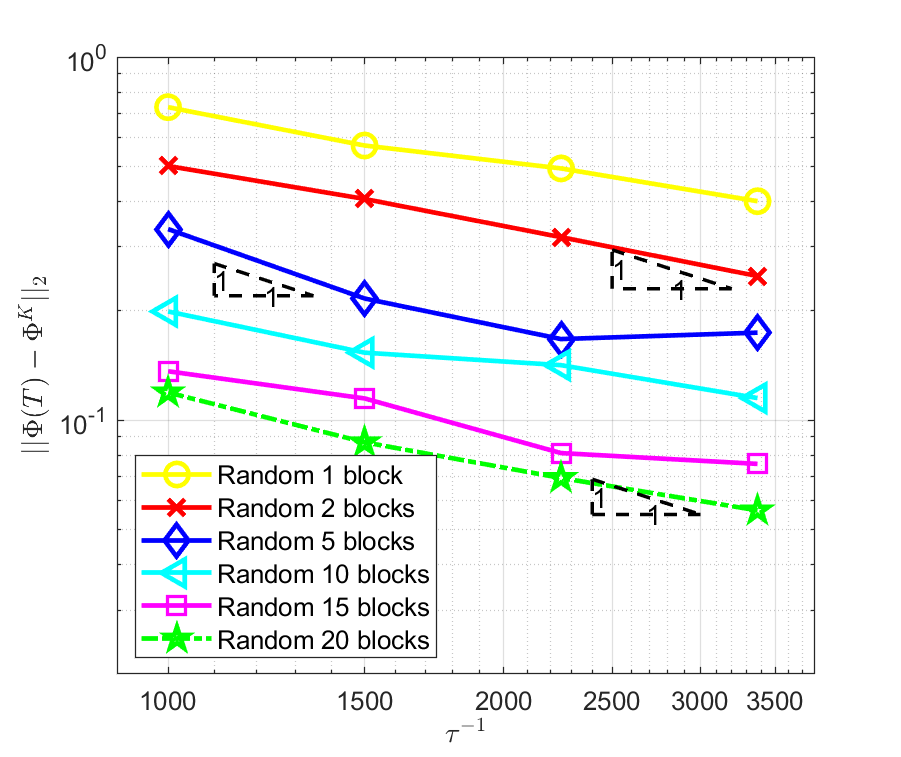}}
	\caption{(Example 1) Error decay for different numerical schemes.}
	\label{1derror}
\end{figure}

\begin{figure}[htb!]
	\centering  
	\subfigure{
		\includegraphics[height=6.0cm]{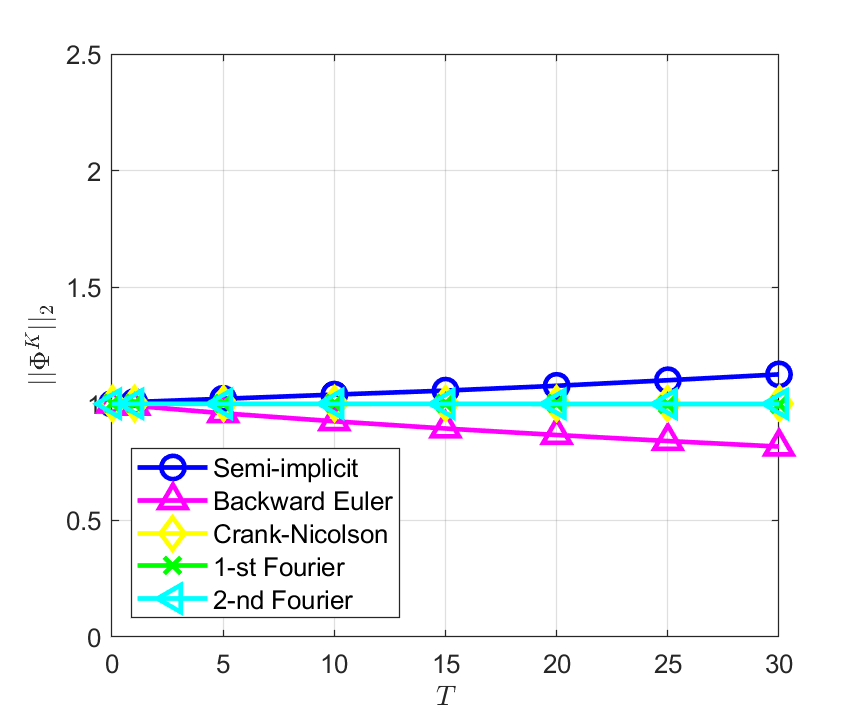}}
	\subfigure{
		\includegraphics[height=6.0cm]{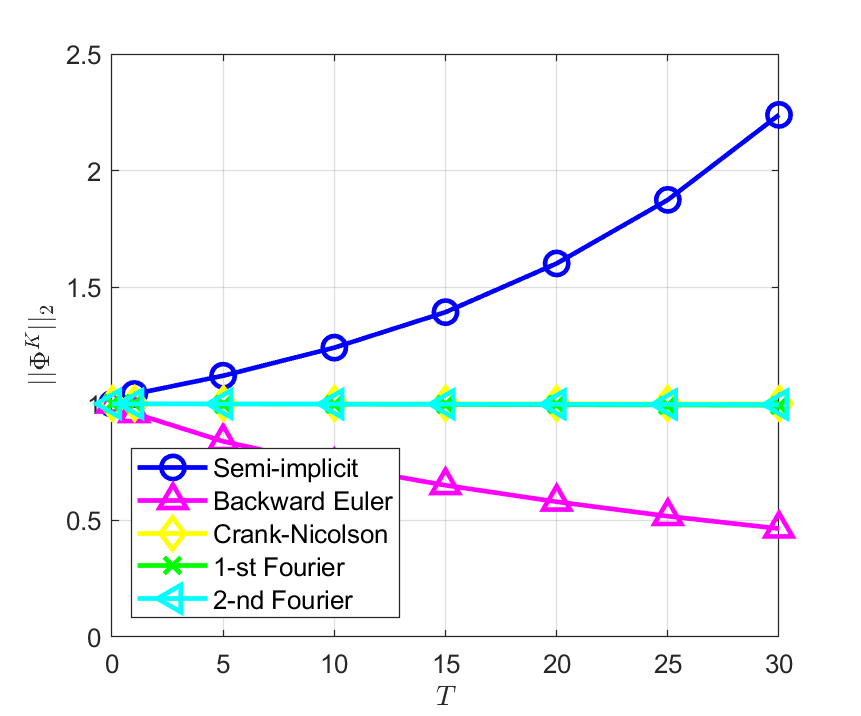}}
	\caption{(Example 1) Mass conservation for different numerical schemes. Left: $\tau=0.01$, Right: $\tau=0.05$.}
	\label{1dnorm}
\end{figure}

The decay of numerical errors with respect to the time step for different splitting methods are presented in Figure \ref{1derror}, from which we see that the convergence rates are consistent with the theoretical predictions.
Compared with the standard splitting methods, the random batch methods possess the same convergence rates, which however have larger pre-constants.
We observe better convergence rates with lager sampling sizes in the figure.
It will be significant to justify how the constants depend on the sampling sizes, which may be discussed in our future work.
We further show the mass evolution for different schemes in Figure \ref{1dnorm}, and observe that the Crank-Nicolson based operator splitting method and Fourier transform based operator splitting method can conserve the mass, while the other schemes can not. This is also consistent with our analysis.

{\bf Example 2.}
(2D incommensurate systems)
Consider a two dimensional incommensurate system obtained by stacking two square lattices together, in which one layer is rotated by an angle $\theta=\pi/10$ with respect to the other. 
More precisely, we take $\mathcal{R}_1=L\cdot A_1\mathbb{Z}^2$ and $\mathcal{R}_2=L\cdot A_2\mathbb{Z}^2$ with
\begin{equation*}
	A_1=\left[ 
	\begin{gathered}
	\begin{matrix}
	1 &0\\0 & 1
	\end{matrix}
	\end{gathered}\right] ,
	\qquad \qquad
	A_2=\left[ 
	\begin{gathered}
	\begin{matrix}
	\cos(\theta) &\cos(\theta+\frac{\pi}{2})  \\\sin(\theta) & \sin(\theta+\frac{\pi}{2}) 
	\end{matrix}
	\end{gathered}\right]	 		 
\end{equation*}
and the lattice constant $L=2$. 
We solve the following Schr\"{o}dinger equation
\begin{equation*}
	\displaystyle i\frac{\partial}{\partial t}\psi(t,x) =\Big(-\frac{1}{2}\Delta+v_1(x)+v_2(x)\Big)\psi(t,x)
\end{equation*}
by taking $\Ec=60$ and $\gamma=5$, where $x\in\R^2$.
The periodic $v_1(x)$ and $v_2(x)$ in this example are given by
\begin{equation}
    v_1(x)=v(A_1x), \quad  v_2(x)=v(A_2x),   \quad \text{with}\ v(x)=\cos\Big(\frac{\pi}{L}x_1\Big)^2\cos\Big(\frac{\pi}{L}x_2\Big)^2
\end{equation}

\begin{figure}[htb!]
\centering
\subfigure{
	\includegraphics[height=6.0cm]{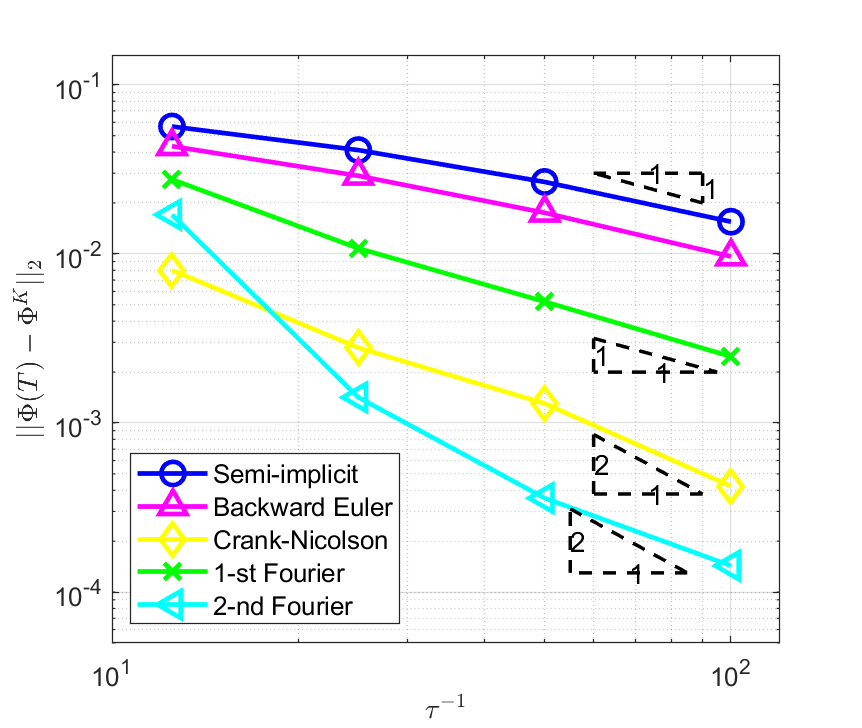}}
\subfigure{
	\includegraphics[height=6.0cm]{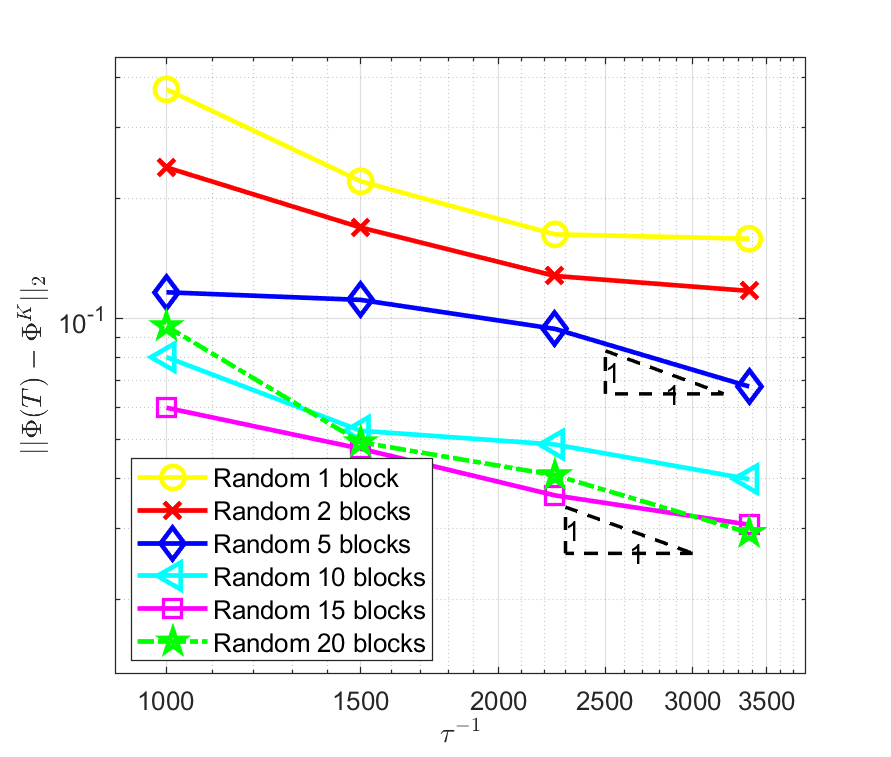}}
\caption{(Example 2) Error decay for different numerical schemes.}
\label{2derror}
\end{figure}

\begin{figure}[htb!]
\centering
\subfigure{
    \includegraphics[height=6.0cm]{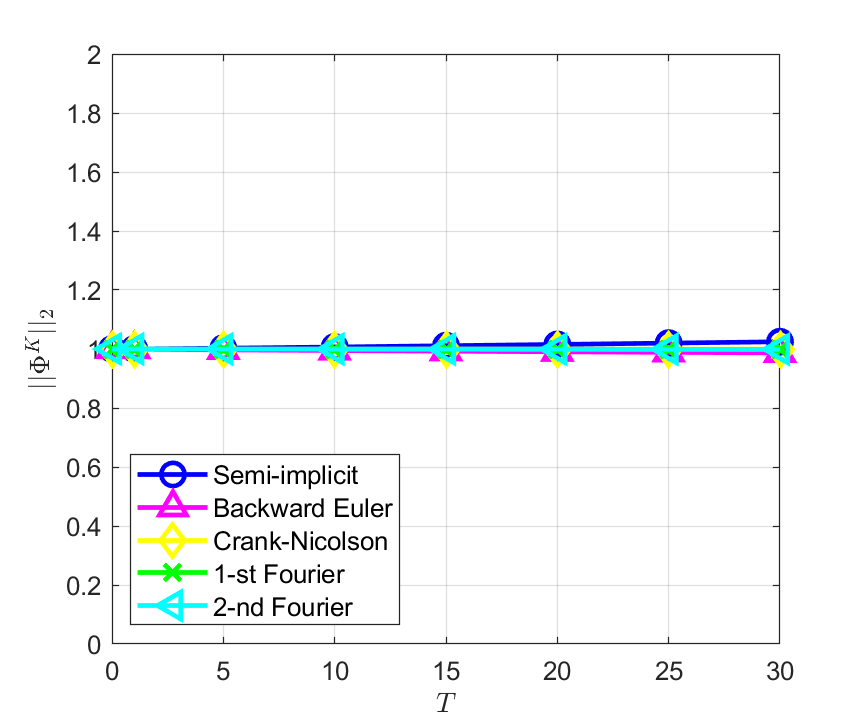}}
\subfigure{
	\includegraphics[height=6.0cm]{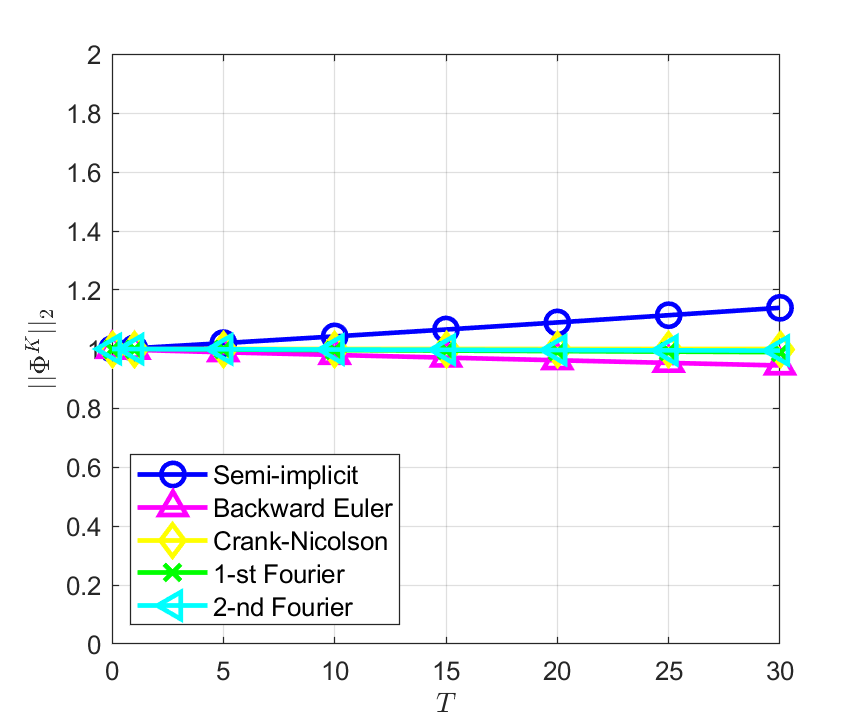}}
\caption{(Example 2) Mass conservation for different numerical schemes. Left: $\tau=0.01$, Right: $\tau=0.05$.}
\label{2dnorm}
\end{figure}

The convergence of numerical errors and conservation of the mass are shown in Figure \ref{2derror} and Figure \ref{2dnorm} respectively, which are consistent with the analysis in this paper.
For the random batch methods, we again observe better convergence with larger sampling size, which is not against our intuition though deserves to be investigated by serious analysis.

{\bf Example 3.}
We consider in this example a two dimensional system with periodic structure in one direction, and incommensurate structure in the other.
Consider the following Schr\"{o}dinger equation
\begin{equation}
	\displaystyle i\frac{\partial}{\partial t}\psi(t,x,y) =\Big(-\frac{1}{2}\Delta+s_1\cos(\beta x)^2\cos(\frac{\pi}{2}y)^2+s_2\cos( x)^2\cos(\frac{\pi}{2} y)^2\Big)\psi(t,x,y)
\end{equation}
with $s_1,\ s_2$ and $\beta$ the parameters that vary in the following experiments.
We take the energy cutoff $\Ec=100$ in this example and the initial state in the following form
\begin{equation}
	\psi_{\Ec}^0(x,y)= \sum_{(m,n)\in \Index{}}\sum_{\{p \in \Z: |G_{3p}|^2\leq 2\Ec\}} 
     e^{-\gamma_x|G_{1m}+G_{2n}|^2}e^{-\gamma_y|G_{3p}|^2}
	e^{i(G_{1m}+G_{2n})x}e^{iG_{3p}y}.
\end{equation} 

\begin{figure}[htb!]
	\centering 
	\subfigure{
	\includegraphics[height=6.0cm]{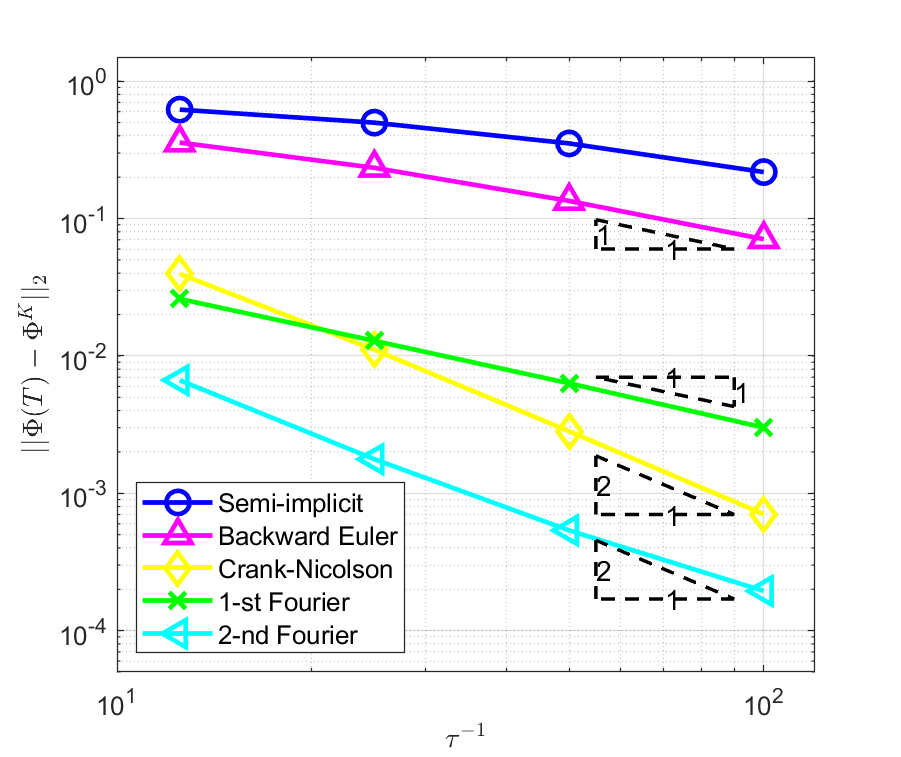}}
	\subfigure{
	\includegraphics[height=6.0cm]{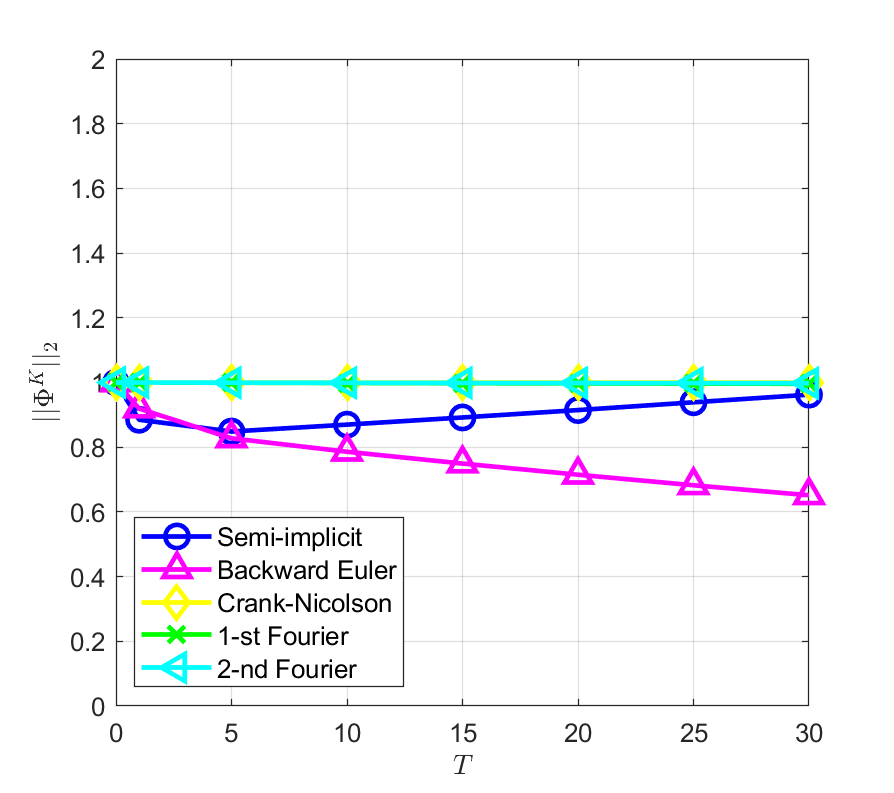}}
	
	\caption{(Example 3) Error decay and mass conservation for different numerical schemes.}
	\label{quasi1D}
\end{figure}

We first take $s_1=1,\ s_2=2$, $\beta=\frac{\sqrt{5}-1}{2}$, $\gamma_x=5$ and $\gamma_y=0.1$ to test the numerical schemes.
We present the error decay and mass evolution in Figure  \ref{quasi1D},  
from which we observe that the convergence rates and mass conservation match our theoretical prediction very well.

\begin{figure}[htb!]
\centering 
\subfigure[$s_1=1,s_2=2,\beta=(\sqrt{5}-1)/2,\gamma_x=5,\gamma_y=0.1$]{
	\label{quasi1Ds11s22}
	\includegraphics[width=6.5in,height=1.0in]{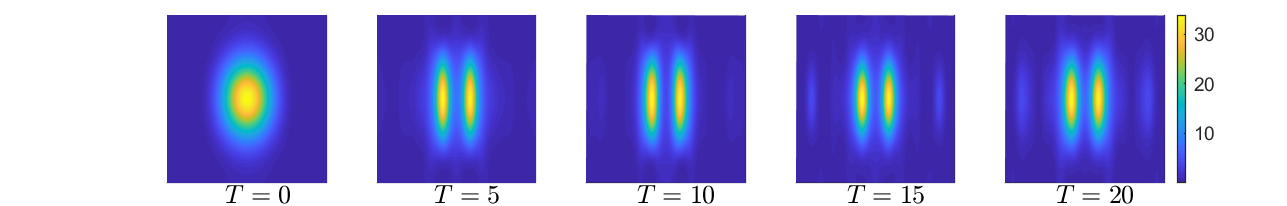}}
	\subfigure[$s_1=0.16,s_2=8,\beta=(\sqrt{5}-1)/2,\gamma_x=5,\gamma_y=0.1$]{
	\label{quasi1Ds1016s28}
	\includegraphics[width=6.5in,height=1.0in]{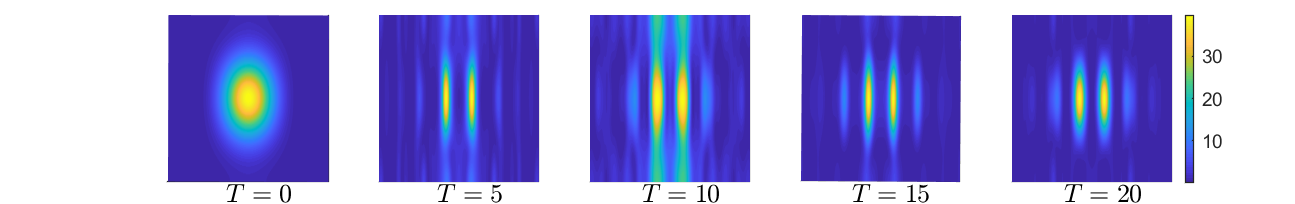}}
    \quad
	\subfigure[$s_1=1,s_2=1,\beta=(1.197+\sqrt{5})/10000,\gamma_x=2,\gamma_y=0.1$]{
	\label{quasi1Dbeta11972s11s22}
	\includegraphics[width=6.5in,height=1.0in]{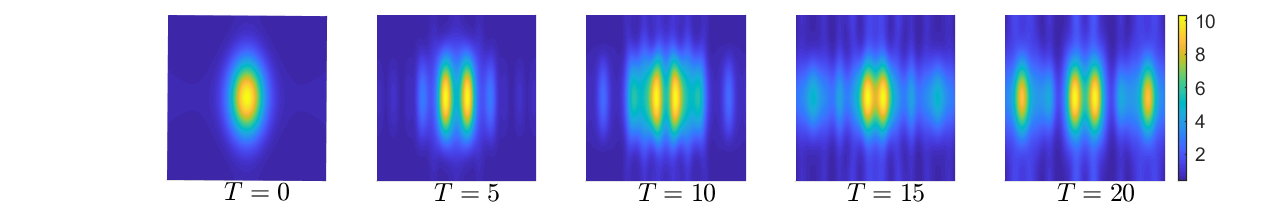}}
	\quad
	\subfigure[$s_1=9,s_2=0.1,\beta=(1.197+\sqrt{5})/10000,\gamma_x=2,\gamma_y=0.1$]{
	\label{quasi1Dbeta11972s18s2002}
   \includegraphics[width=6.5in,height=1.0in]{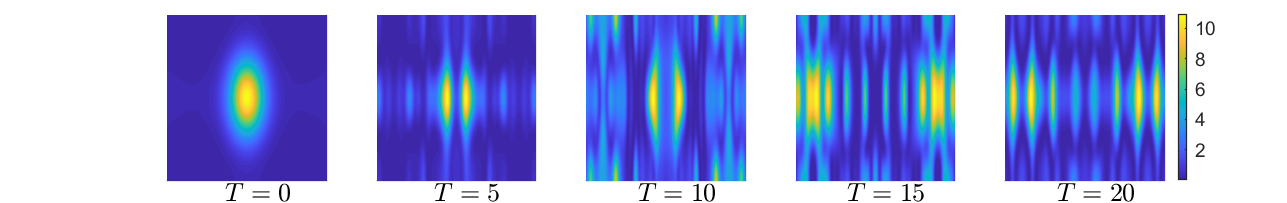}}
\caption{The evolution of the wave functions in different incommensurate systems.}
\label{quasi1Devolution}
\end{figure}

An interesting phenomena in the incommensurate systems is the localization of quantum states, see e.g. \cite{anderson1958absence,chen2020plane,roati2008anderson}.
We simulate the localized-to-extended transition by artificially controlling the potential strengths $s_1,~s_2$ and ratios $\beta$ of the incommensurate systems, following the experiments in \cite{roati2008anderson}. 
We present the evolution of square norm of the wave function $|\psi_{\Ec}(x,y)|^2$ in Figure \ref{quasi1Devolution}, and observe different behaviours of the evolution, which shows a cross-over from localized to extended states in the $x$-direction.

\section{Conclusions}
\label{sec:conclusion}

In this paper, we propose some temporal schemes for linear time-dependent Schr\"{o}dinger equations of incommensurate systems.
The high dimensional semidiscrete problem from the plane wave spatial discretization is split into low dimension sub-problems corresponding to different periodic layers.
These algorithm are natural to be combined with the stochastic techniques and parallel computing.
The convergence and mass conservation of our algorithm are verified by both theoretical analysis and numerical experiments.
Both algorithms and theories in this paper can be generalized directly to incommensurate systems with more than two layers.
The extension of this method to more sophisticated physical models with nonlinear terms will be investigated in our future work.

\section*{Acknowledgements}
This work was partially supported by National Key R\&D Program of China 2019YFA0709\\
600, 2019YFA0709601. T. Wang and H. Chen’s work was also partially supported by the Natural Science Foundation of China under grant 11971066. A. Zhou’s work was also partially supported by the Key Research Program of Frontier Sciences of the Chinese Academy of Sciences under grant QYZDJ-SSW-SYS010, and the National Science Foundation of China under grant 11671389. Y. Zhou’s work was also partially supported by the National Science Foundation of China under grant 12004047.

\begin{appendices}

\section{Standard operator splitting methods}
\label{sec:appendix splitting method}

The operator splitting method \cite{glowinski2017splitting,marchuk1968some,strang1968construction} can be used in a linear system of ordinary differential equations of the following abstract form. 
Let $S$ denote a vector space, for $u(t):[0,T]\rightarrow S$, $0<T< +\infty$,
\begin{equation}\label{splitODE}
\left\{
\begin{array}{l}
\displaystyle\frac{\mathrm{d}u(t)}{\mathrm{d}t}=Au(t),\quad t\in[0,T],\\[2ex]
u(0)=u_0,
\end{array}
\right.	
\end{equation}	
where operator $A: S\rightarrow S$ has a decomposition
\begin{equation}\label{splitA}
     A=\sum_{j=1}^J A_j
\end{equation}
with $J\geq 2$. 
The operator splitting schemes are designed to take advantage of the decomposition \eqref{splitA} for solving \eqref{splitODE}.
We give the Lie-Trotter and the Strang splitting schemes in the following for the simplest case of $J=2$.

Let $\tau>0$ be the time step, and $t_k=k\tau,\ k=1,2,\dots,K$, $T=K\tau$. 
The Lie-Trotter splitting scheme reads as follows:
Let $u^{(0)}_2(0)=u_0$ and
\begin{equation*}\label{Liesplitting1}
\left\{
\begin{array}{l}
\displaystyle \frac{\mathrm{d}u^{(k)}_1(t)}{\mathrm{d}t}=A_1u^{(k)}_1(t), \quad  t\in (t_{k-1},t_k], \qquad {\rm with} \quad u_1^{(k)}(t_{k-1})=u_2^{(k-1)}(t_{k-1}), \\[2ex]
\displaystyle \frac{\mathrm{d}u^{(k)}_2(t)}{\mathrm{d}t}=A_2u^{(k)}_2(t),\quad t\in (t_{k-1},t_k], \qquad {\rm with} \quad
u_2^{(k)}(t_{k-1})=u_1^{(k)}(t_k)
\end{array}
\right.	
\end{equation*}
for $k=1,2,\dots,K$.
The Strang splitting scheme \cite{strang1968construction} reads as follows (with $t^{k-1/2}=(k-1/2)\tau$):
Let $u^{(0)}_3(0)=u_0$ and
\begin{equation*}\label{Strangsplitting1}
\left\{
\begin{array}{l}
\displaystyle \frac{\mathrm{d}u^{(k)}_1(t)}{\mathrm{d}t}=A_1u^{(k)}_1(t), \quad  t\in (t_{k-1},t_{k-1/2}], \quad {\rm with} \quad u_1^{(k)}(t_{k-1})=u_3^{(k-1)}(t_{k-1}), \\[2ex]
\displaystyle \frac{\mathrm{d}u^{(k)}_2(t)}{\mathrm{d}t}=A_2u^{(k)}_2(t),\quad t\in (t_{k-1},t_k], \ \ \qquad {\rm with} \quad
u_2^{(k)}(t_{k-1})=u_1^{(k)}(t_{k-1/2}), \\[2ex]
\displaystyle \frac{\mathrm{d}u^{(k)}_3(t)}{\mathrm{d}t}=A_1u^{(k)}_3(t), \quad  t\in (t_{k-1/2},t_k], \qquad {\rm with} \quad
u_3^{(k)}(t_{k-1/2})=u_2^{(k)}(t_k),
\end{array}
\right.	
\end{equation*}
for $k=1,2,\dots,K$.

At time interval $[t_k,t_{k+1}]$,
operator splitting approximation is usually given by 
\begin{equation*}\label{liesplitting}
\ \quad u(t_{k+1})\approx e^{A_2 \tau}e^{A_1 \tau}u(t_k)
\end{equation*}
for the Lie-Trotter splitting scheme, and 
\begin{equation*}\label{strangsplitting}
\ \quad u(t_{k+1})\approx e^{A_1 \tau/2}e^{A_2 \tau}e^{A_1 \tau/2}u(t_k)
\end{equation*}
for the Strang splitting scheme.
From the Taylor expansion and the Baker-Campbell-Hausdorff formula \cite{hairer2006geometric} , it is obvious to see that the approximation error of the Lie-Trotter splitting is of first order $\mathcal{O}(\tau)$, and the error of the Strang splitting is of second order $\mathcal{O}(\tau^2)$. We note that splitting approximations of higher order accuracy can be constructed \cite{hairer2006geometric,yoshida1990construction}.

\section{Proofs of convergence and mass conservation}
\label{sec:appendix proofs}

Let $\vecn{}(t)$ be the solution of \eqref{ODE}, $t\in [0,T]$, and $\vecn{k}$ be the solution obtained by different numerical methods. 
Denote by $\rho(H)$ the spectral radius of the matrix $H$ in the Appendix. 

\vskip 0.3cm
\noindent\textbf{Proof of Theorem \ref{error_semi}.}
The numerical scheme \eqref{semi-implicitV1V2} can be formulated as
	\begin{eqnarray} \label{SInumerical}
	\vecn{k}=\big(I+ i\tau  D\big)^{-1}\big(I-i\tau (V^1+V^2)\big)\vecn {k-1} \qquad  \text{for} \ k=1,2,\dots,K,
	\end{eqnarray}
where $D,V^1,V^2$ have the formula (\ref{DV1V2}), and 
	\begin{equation*}
	\big(I-i\tau (V^1+V^2)\big)_{mn,m'n'}=\delta_{mm'}\delta_{nn'}-i\tau \vm{(m-m')}\delta_{nn'}-i\tau \vn{(n-n')}\delta_{mm'}.
	\end{equation*}
Since $v_1\in L^2_{\#,1}(\Gamma_1)$ and $v_2\in L^2_{\#,2}(\Gamma_2)$, we have
\begin{equation*}
 \displaystyle ||v_1||_{L^2_{\#,1}}^2=\sum_{\{m\in \Z^d:G_{1m}\in \RL_1^*\}}|\vm{m}|\leq C_1 \qquad \text{and} \qquad  ||v_2||_{L^2_{\#,2}}^2=\sum_{\{n\in \Z^d:G_{2n}\in \RL_2^*\}}|\vn{n}|\leq C_2.
\end{equation*}
From \eqref{DV1V2}, $V^1$ and $V^2$ are Hermitian matrices, $(V^1+V^2)^*(V^1+V^2)$ is a positive semi-definite matrix, thus all of its eigenvalues are non-negative.
It can be shown that
\begin{equation*}
\begin{split}
 \rho^2\big(I & -i\tau (V^1+V^2)\big)= \lambda_{\max}\big( (I-i\tau (V^1+V^2))^*(I-i\tau (V^1+V^2))\big)\\
  &=\lambda_{\max}\big( I+\tau^2 (V^1+V^2)^*(V^1+V^2)\big)\\
  &\leq 1+\tau^2 \big(|\Index{}|(C_1+C_2)\big) \\
\end{split}   
\end{equation*}	
Then we get
\begin{equation}
\rho\big(I -i\tau (V^1+V^2)\big)\leq \sqrt{1+\tau^2 \big(|\Index{}|(C_1+C_2)\big)} \leq 1+E\tau,    
\end{equation}
where $E=\sqrt{|\Index{}|(C_1+C_2)}>0$ is a constant depending on the energy cutoff $E_c$. Moreover, the spectral radius 
$\rho\big((I+i\tau D)^{-1}\big)=1$, then the spectral radius of the amplification factor corresponding to \eqref{SInumerical} is given by 
	\begin{equation} \label{SIspectralradius}
	\rho\big(\big(I+i\tau D\big)^{-1}\big(I-i\tau (V^1+V^2)\big)\big)\leq 	\rho\big(\big(I+i\tau D\big)^{-1}\big) \rho\big(I-i\tau (V^1+V^2)\big)\leq 1+E\tau.
	\end{equation}
	Using \eqref{SInumerical} and \eqref{SIspectralradius}, we obtain 
	\begin{align}\label{SIstability}
	 ||\vecn{k}||_{\ell_2}&=\Big|\Big|\big(I+ i\tau  D\big)^{-1}\big(I-i\tau (V^1+V^2)\big) \vecn {k-1}\Big|\Big|_{\ell_2}\nonumber\\	
	&\leq \rho\big(\big(I+ i\tau  D\big)^{-1}\big(I-i\tau (V^1+V^2)\big)\big)||\vecn {k-1}||_{\ell_2}\leq (1+E\tau)||\vecn {k-1}||_{\ell_2}.
\end{align}
Let
\begin{equation*}
  {\omega} ^k = \big(I+ i\tau  D\big)^{-1}\big(I-i\tau (V^1+V^2)\big)\vecn {}(t_{k-1}),
\end{equation*} 
then it can be shown that
\begin{equation}\label{SIlocalerror}
||\vecn{}(t_k) - {\omega} ^k ||_{\ell_2}\leq C\tau^2.
\end{equation}
From \eqref{SIstability}, we have
\begin{equation*}
||{\omega}^k-\vecn{k}||_{\ell_2} \leq (1+E\tau) ||\vecn{}(t_{k-1})-\vecn{k-1}||_{\ell_2}. 
\end{equation*}
Applying \eqref{SInumerical}, \eqref{SIstability} and \eqref{SIlocalerror}, we can obtain by triangle inequality and induction that 
\begin{equation*}\label{exact_numerical_SI}
\begin{split}
&||\vecn{}(t_k)-\vecn{k}||_{\ell_2}\leq
||\vecn{}(t_k)-{\omega}^k||_{\ell_2}+||{\omega}^k-\vecn{k}||_{\ell_2} \\
&\leq C\tau^2 +(1+E\tau)||\vecn{}(t_{k-1})-\vecn{k-1}||_{\ell_2}
 \leq\frac{e^{ET}-1}{E}\tau=C_T\tau \xqedhere{89pt}{\qed}. 
\end{split} 
\end{equation*}

\vskip 0.3cm
\noindent\textbf{Proof of Theorem \ref{error_splittingbE}.}
For simplicity of notations, we denote by $H=H_1+H_2$, where
\begin{equation*}\label{H1H2}
H_1=\frac{1}{2}D+I_1\otimes \Vc^{1}
\qquad \text{and}\qquad 
H_2=\frac{1}{2}D+\Vc^{2}\otimes I_2.
\end{equation*}
Since $H_1$ and $H_2$ are Hermitian matrices, they can be diagonalized by a unitary matrix, and that the resulting diagonal matrix has only real entries. 
Denote by $\tilde{\Phi}^k$ the solution obtained by the Lie-Trotter splitting method at $t_k$, we have
\begin{equation}\label{stability_lie}
    ||\tilde{\Phi}^k||_{\ell_2}=||e^{-i\tau H_2}\tilde{\Phi}^{k-1/2}||_{\ell_2}=||\tilde{\Phi}^{k-1/2}||_{\ell_2}=||e^{-i\tau H_1}\tilde{\Phi}^{k-1}||_{\ell_2}=||\tilde{\Phi}^{k-1}||_{\ell_2}.
\end{equation}
Moreover, the amplification factors corresponding to \eqref{tsbEtimestepping1} and \eqref{tsbEtimestepping2} are given by
\begin{equation*}
\rho\big((I+i\tau H_1)^{-1}\big)\leq 1\qquad \text{and}
\qquad  \rho\big((I+i\tau H_2)^{-1}\big)\leq 1.
\end{equation*}
We thus obtain
\begin{equation}\label{stability_tsbE}
	\begin{split}
	||\vecn{k}||_{\ell_2}&\leq \rho\big((I+i\tau H_2)^{-1}(I+i\tau H_1)^{-1}\big)||\vecn {k-1}||_{\ell_2} \leq||\vecn {k-1}||_{\ell_2}.
	\end{split}
\end{equation}
Let
\begin{equation*}
{\omega} ^k = e^{-i\tau H_2}e^{-i\tau H_1}\vecn{}(t_{k-1})\qquad \text{and} \qquad \mu^k=\big(I+i \tau H_2\big)^{-1}\big(I+i \tau H_1\big)^{-1}\tilde{\Phi}^{k-1},
\end{equation*} 
then it can be shown that
\begin{equation}\label{localbEandsplitting}
||\vecn{}(t_k) - {\omega} ^k ||_{\ell_2}\leq C\tau^2\qquad\text{and} \qquad || \mu ^k - \tilde{\Phi} ^k||_{\ell_2}\leq C\tau^2.
\end{equation}
From \eqref{stability_lie} and \eqref{stability_tsbE}, we have
\begin{equation}\label{stablebEandsplitting}
||\omega^k-\tilde{\Phi}^k||_{\ell_2} = ||\vecn{}(t_{k-1})-\tilde{\Phi}^{k-1}||_{\ell_2}   
\qquad {\rm and}\qquad 
 ||\mu^k -\vecn k||_{\ell_2}\leq ||\tilde{\Phi}^{k-1}-\vecn{k-1}||_{\ell_2}.   
\end{equation}
Using \eqref{localbEandsplitting} and \eqref{stablebEandsplitting}, we can obtain by the triangle inequality and induction that
\begin{equation*}\label{exact_numerical_bE}
\begin{split}
 ||\vecn{}(t_k)-\vecn{k}||_{\ell_2}
&\leq ||\vecn{}(t_k) - {\omega} ^k +{\omega} ^k-\tilde{\Phi}^k||_{\ell_2} + ||\tilde{\Phi}^k -\mu ^k + \mu ^k - \vecn{k} ||_{\ell_2}\\
& \leq C \tau^2+ ||\vecn{}(t_{k-1})-\tilde{\Phi}^{k-1}||_{\ell_2}
+||\tilde{\Phi}^{k-1}-\vecn{k-1}||_{\ell_2}\leq C_T\tau.  
\xqedhere{52pt}{\qed}
\end{split}
\end{equation*}

\vskip0.3cm
\noindent\textbf{Proof of Theorem \ref{error_splittingCN}.}
For simplicity of notations, we denote by $H=H_1+H_2$, where
\begin{equation*}
H_1=\frac{1}{2}D+I_1\otimes \Vc^{1}
\qquad \text{and}\qquad  
H_2=\frac{1}{2}D+\Vc^{2}\otimes I_2.
\end{equation*}
Denote by $\tilde{\Phi}^{k}$ the solution obtained by the Strang splitting method at $t_k$. Since $H_1$ and $H_2$ are Hermitian matrices, we have
\begin{align}\label{stableStrang}
    ||\tilde{\Phi}^{k}||_{\ell_2} &=||e^{-i\frac{\tau}{2} H_1}\tilde{\Phi}^{k-1/3}||_{\ell_2}=||\tilde{\Phi}^{k-1/3}||_{\ell_2}
    =||e^{-i\tau H_2}\tilde{\Phi}^{k-2/3}||_{\ell_2} \nonumber \\
   & =||\tilde{\Phi}^{k-2/3}||_{\ell_2}=||e^{-i\frac{\tau}{2}H_1}\tilde{\Phi}^{k-1}||_{\ell_2}=||\tilde{\Phi}^{k-1}||_{\ell_2}.
\end{align}
Multiplying $(\vecn{k-2/3}+\vecn {k-1})^*$ to the first equation of \eqref{tsCN} at time interval $[t_{k-1},t_k]$, we obtain
\begin{equation}\label{stableCN1}
i \ \frac{2}{\tau}\Big(\vecn{k-2/3}+\vecn {k-1}\Big)^*\Big(\vecn{k-2/3}-\vecn {k-1}\Big)=\frac{1}{2}\Big(\vecn{k-2/3}+\vecn {k-1}\Big)^*H_1\Big(\vecn{k-2/3}+\vecn {k-1}\Big).
\end{equation}
Since $H_1$ is a Hermitian matrix, the right of \eqref{stableCN1} is real. Taking the imaginary part of \eqref{stableCN1}, we obtain
\begin{equation*}
||\vecn{k-2/3}||^2_{\ell_2}=||\vecn{k-1}||^2_{\ell_2}.
\end{equation*} 
Similarly, we have
\begin{equation*}
||\vecn{k-1/3}||_{\ell_2}^2=||\vecn{k-2/3}||_{\ell_2}^2 \qquad \text{and} \qquad ||\vecn{k}||_{\ell_2}^2=||\vecn{k-1/3}||_{\ell_2}^2.
\end{equation*} 
Then we obtain unconditional stability of numerical scheme \eqref{tsCN}
\begin{equation}\label{stabilityCN}
||\vecn{k}||_{\ell_2}^2=||\vecn{k-1}||_{\ell_2}^2.
\end{equation}
Let
\begin{equation*}
{\omega} ^k = e^{-i\tau/2 H_1} e^{-i\tau H_2}e^{-i\tau/2 H_1}\vecn{}(t_{k-1})\qquad \text{and}  \qquad
\mu^{k}=\Sc_1(\tau)\Sc_2(\tau)\Sc_1(\tau)\tilde{\Phi}^{k-1},
\end{equation*}
where $\Sc_1(\tau)$ and $\Sc_2(\tau)$ are defined by 
\begin{equation*}
 \Sc_1(\tau) =  \big(4I+i \frac{\tau}{2} H_1\big)^{-1}\big(4I-i \frac{\tau}{2} H_1\big) \qquad {\rm and}\qquad 
 \Sc_2(\tau)=\big(2I+i \frac{\tau}{2} H_2\big)^{-1}\big(2I-i\frac{\tau}{2} H_2\big),
\end{equation*}
then it can be shown that 
\begin{equation}\label{localCNandsplitting}
||\vecn{}(t_k) - {\omega} ^k ||_{\ell_2}\leq C\tau^3\qquad\text{and} \qquad || \mu ^k - \tilde{\Phi} ^k||_{\ell_2}\leq C\tau^3.
\end{equation}
By \eqref{stableStrang} and \eqref{stabilityCN}, we have
\begin{equation}\label{stableCNandsplitting}
||\omega^k-\tilde{\Phi}^k||_{\ell_2} = ||\vecn{}(t_{k-1})-\vecn{k-1}||_{\ell_2}   
\qquad {\rm and}\qquad 
 ||\mu^k -\vecn k||_{\ell_2} =  ||\tilde{\Phi}^{k-1}-\vecn{k-1}||_{\ell_2}.   
\end{equation}
Using \eqref{localCNandsplitting} and \eqref{stableCNandsplitting}, we can show by triangle inequality and induction that
\begin{equation*}\label{exact_numerical_CN}
\begin{split}
 ||\vecn{}(t_k)-\vecn{k}||_{\ell_2}
&\leq ||\vecn{}(t_k) - {\omega} ^k +{\omega} ^k-\tilde{\Phi}^k||_{\ell_2} + ||\tilde{\Phi}^k- \mu ^k + \mu ^k - \vecn{k} ||_{\ell_2}\\
& \leq C \tau^3+ ||\vecn{}(t_{k-1})-\tilde{\Phi}^{k-1}||_{\ell_2}
+||\tilde{\Phi}^{k-1}-\vecn{k-1}||_{\ell_2}\leq C_T\tau^2.
\xqedhere{50pt}{\qed}
\end{split}
\end{equation*}

\vskip0.3cm
\noindent\textbf{Proof of Theorem \ref{massconservationtsCN}.}
From \eqref{wavefunctionapprox}, we have
\begin{equation*}
 \psi_{\Ec}^k= \sum_{(m,n)\in \Index{} }\phi_{mn}^ke^{i(G_{1m}+G_{2n})\cdot x}.
\end{equation*}
Using the orthogonal condition \eqref{ortho}, we get
\begin{equation*}
N(\psi_{\Ec}^k)=\sum_{(m,n)\in \Index{}}|\phi_{mn}^k|^2=||\vecn{k}||_{\ell_2}^2 \qquad \text{for}\ k\geq 0.
\end{equation*}
Then \eqref{mass conservation tsCN} is equivalent to
\begin{equation}\label{normphi}
||\vecn{k}||_{\ell_2}^2=||\vecn{0}||_{\ell_2}^2 \qquad  \text{for}\ k\geq 0.
\end{equation} 
Multiplying $(\vecn{k+1/3}+\vecn k)^*$ to the first equation of \eqref{tsCN}, we obtain
\begin{equation}\label{mass}
i \ \frac{2}{\tau}\Big(\vecn{k+1/3}+\vecn k\Big)^*\Big(\vecn{k+1/3}-\vecn k\Big)=\frac{1}{2}\Big(\vecn{k+1/3}+\vecn k\Big)^*H_1\Big(\vecn{k+1/3}+\vecn k\Big).
\end{equation}
Since $H_1$ is a Hermitian matrix, the right of \eqref{mass} is real. Taking the imaginary part of \eqref{mass}, we obtain
\begin{equation}\label{normek}
||\vecn{k+1/3}||^2_{\ell_2}=||\vecn{k}||^2_{\ell_2}.
\end{equation} 
Similarly, we have
\begin{equation}\label{normek+1}
||\vecn{k+2/3}||_{\ell_2}^2=||\vecn{k+1/3}||_{\ell_2}^2 \qquad \text{and} \qquad ||\vecn{k+1}||_{\ell_2}^2=||\vecn{k+2/3}||_{\ell_2}^2.
\end{equation} 
Combining \eqref{normek+1} with \eqref{normek}, we get
\begin{equation*}
||\vecn{k+1}||_{\ell_2}^2=||\vecn{k}||_{\ell_2}^2,
\end{equation*} 
which implies that \eqref{normphi}. And thus the mass conservation in \eqref{mass conservation tsCN} holds for $ k \geq 0 $.
\hfill $\square$

\vskip0.3cm
\noindent\textbf{Proof of Theorem \ref{error_splittingtsFourier}.}
Since $V^1$ and $V^2$ are Hermitian matrices, and $D$ is a diagonal matrix, we have 
\begin{equation*}\label{stabilitysplittingFourier}
\begin{split}
\big|\big|\vecn{k}\big|\big|_{\ell_2} &=\big|\big|e^{-i\tau V^2}\vecn{k-1/3}\big|\big|_{\ell_2}=\big|\big|\vecn{k-1/3}\big|\big|_{\ell_2}
=\big|\big|e^{-i\tau V^1}\vecn{k-2/3}\big|\big|_{\ell_2}\\
&=\big|\big|\vecn{k-2/3}\big|\big|_{\ell_2}
=\big|\big|e^{-iD}\vecn{k-1}\big|\big|_{\ell_2} 
=\big|\big|\vecn{k-1}\big|\big|_{\ell_2}.
\end{split}
\end{equation*}
Let 
\begin{equation*}
{\omega}^k =e^{-i\tau V^2}e^{-i\tau V^1}e^{-iD}\vecn{}(t_{k-1}),  
\end{equation*}
it can be shown that 
\begin{equation}\label{locallietsF}
||\vecn{}(t_k) - \omega^k||_{\ell_2} \leq C \tau^2
\end{equation}
and
\begin{align}\label{stabilitytsF}
||\omega^k-\vecn k||_{\ell_2} &= ||e^{-i\tau V^2}e^{-i\tau V^1}e^{-iD}\vecn{}(t_{k-1})-e^{-i\tau V^2}e^{-i\tau V^1}e^{-iD}\vecn{k-1}||_{\ell_2}\nonumber \\
&= ||\vecn{}(t_{k-1})-\vecn{k-1}||_{\ell_2}.
\end{align}
From \eqref{locallietsF} and \eqref{stabilitytsF},
we obtain the following convergence rate by triangle inequality and induction
\begin{equation*}
 ||\vecn{}(t_k)-\vecn k||_{\ell_2}=||\vecn{}(t_k)-\omega^k +\omega^k-\vecn k||_{\ell_2}
 \leq C\tau^2
+ ||\vecn{}(t_{k-1})-\vecn{k-1}||_{\ell_2}
\leq C_T\tau.
\xqedhere{28pt}{\qed}
\end{equation*}

\vskip0.3cm
\noindent\textbf{Proof of Theorem \ref{massconservationtsFourier}.}
Using orthogonal condition \eqref{ortho}, iFFT and FFT give
\begin{equation}\label{normfft}
\sum_{m\in\Index1}|\psi_{mn}^{k+1/3}|^2 
=|\Index1| \sum_{m\in\Index1}|\phi_{mn}^{k+1/3}|^2 \quad {\rm and} \quad 
\sum_{m\in\Index1}|\phi_{mn}^{k+1}|^2 
=\frac{1}{|\Index1|} \sum_{m\in\Index1}|\psi_{mn}^{k+1}|^2
\end{equation}
for $n\in \Index2$. From Algorithm \ref{alg:tsF}, one has
\begin{equation}\label{lapterm}
 ||\vecn{k+1/3}||_{\ell_2}^2 =\sum_{m\in \Index{1}}\sum_{n\in\Index{2}}|\exp(-i\tau |G_{1m}+G_{2n}|^2/2)\phi_{mn}^{k}|^2=||\vecn{k}||_{\ell_2}^2. 
\end{equation}
Using \eqref{normfft} and \eqref{lapterm}, we get
\begin{equation*}
\begin{split}
||\vecn{k+1}||_{\ell_2}^2&=\sum_{m\in \Index{1}}\sum_{n\in\Index{2}}|\phi_{mn}^{k+1}|^2=
\sum_{m\in \Index1}\sum_{n\in\Index2}\frac{1}{|\Index1|}|\psi_{mn}^{k+1}|^2 \\
& = \sum_{m \in \Index1}\sum_{n\in\Index2}\frac{1}{|\Index1|} |\exp(-i\tau v_n^{(2)})\psi_{mn}^{k+2/3}|^2 = \sum_{m\in \Index1}\sum_{n\in\Index2}\frac{1}{|\Index1|}|\psi_{mn}^{k+2/3}|^2\\
& = \sum_{n\in\Index2} \sum_{m\in \Index1}\frac{1}{|\Index1|} |\exp(-i\tau v_m^{(1)})\psi_{mn}^{k+1/3}|^2 = \sum_{n\in\Index2} \sum_{m\in \Index1}\frac{1}{|\Index1|}|\psi_{mn}^{k+1/3}|^2 \\
&=\sum_{n\in\Index2} \sum_{m\in \Index1}|\phi_{mn}^{k+1/3}|^2 
= ||\vecn {k+1/3}||_{\ell_2}^2 = ||\vecn k||_{\ell_2}^2,
\end{split}
\end{equation*}
which implies that \eqref{normphi}.
And thus the mass conservation in \eqref{masstsF} holds for $ k \geq 0 $.
\hfill $\square$

\end{appendices}

\small
\bibliographystyle{plain}
\bibliography{bib.bib}

\end{document}